\documentclass[aps,prb,onecolumn,preprint, longbibliography,superscriptaddress]{revtex4-2}
\usepackage{adjustbox}
\usepackage{rotating}
\usepackage{amsmath,amssymb,amsfonts,upgreek}
\usepackage{color}
\usepackage{graphicx}
\usepackage{setspace}
\usepackage{bm}
\usepackage{hyperref}
\usepackage{adjustbox}
\usepackage{dsfont}
\usepackage{braket}
\usepackage{multirow}
\usepackage{float}
\usepackage[T1]{fontenc}
\newcommand{\cb}{C$_{\text{B}}$}
\newcommand{\cn}{C$_{\text{N}}$}
\newcommand{\vb}{$V_{\text{B}}$}
\newcommand{\vn}{$V_{\text{N}}$} 
\newcommand{\shr}{$S_{\text{HR}}$} 
\newcommand{\efc}{$E_{\text{FC}}$} 

\begin{document}

\title{Dielectric environment sensitivity of carbon centres in hexagonal boron nitride}

\def \Radbound{Institute for Molecules and Materials, Radboud University, Heijendaalseweg 135, 6525 AJ Nijmegen, Netherlands}
\def \Finland{Faculty of Engineering and Natural Sciences, Tampere University, 33720 Tampere, Finland}
\def \NY{Center for Computational Quantum Physics, Flatiron Institute, 162 5$^{th}$ Avenue, New York, NY 10010, USA}
\def \NYY{Department of Physics and Astronomy, Stony Brook University, Stony Brook, New York 11794-3800, USA}
\def \MSE{Department of Materials Science and Engineering, National University of Singapore, 117575, Singapore}
\def \CADM{Centre for Advanced 2D Materials, National University of Singapore, 117546, Singapore} 
\def \IFIM{Institute for Functional Intelligent Materials, National University of Singapore, 117544, Singapore}
\def \CHEM{Department of Chemistry, National University of Singapore, 117543, Singapore}
\def \JAPAN{Research Center for Functional Materials, National Institute for Materials Science, Tsukuba 305-0044, Japan}
\def \JAPANN{International Center for Materials Nanoarchitectonics, National Institute for Materials Science, 305-0044, Japan}
\def \China{Guangxi Key Laboratory of Information Materials, Guilin University of Electronic Technology, Guilin 541004, China}

\def \FUW{Faculty of Physics, University of Warsaw, ul. Pasteura 5, 02-093 Warszawa, Poland}
\def \CENT{CENTERA Labs, Institute of High Pressure Physics, PAS, PL-01-142 Warsaw, Poland}
\def \LNCMI{Laboratoire National des Champs Magn\'etiques Intenses, CNRS-UGA-UPS-INSA-EMFL, 25 Av. des Martyrs, 38042 Grenoble, France}

\title{Dielectric environment sensitivity of carbon centres in hexagonal boron nitride}

\author{Danis~I.~Badrtdinov}
\affiliation{\Radbound}

\author{Carlos~Rodriguez-Fernandez}
\affiliation{\Finland}

\author{Magdalena~Grzeszczyk}
\affiliation{\IFIM}

\author{Zhizhan~Qiu}
\affiliation{\CHEM}

\author{Kristina~Vaklinova}
\affiliation{\IFIM}
\affiliation{\MSE}

\author{Pengru~Huang}
\affiliation{\IFIM}
\affiliation{\MSE}
\affiliation{\China}

\author{Alexander~Hampel}
\affiliation{\NY}

\author{Kenji~Watanabe}
\affiliation{\JAPAN}

\author{Takashi~Taniguchi}
\affiliation{\JAPANN}

\author{Lu~Jiong}
\affiliation{\CHEM}
\affiliation{\CADM}

\author{Marek~Potemski}
\affiliation{\LNCMI}
\affiliation{\CENT}
\affiliation{\FUW}

\author{Cyrus~E.~Dreyer}
\affiliation{\NY}
\affiliation{\NYY}

\author{Maciej~Koperski}
\email{msemaci@nus.edu.sg}
\affiliation{\IFIM}
\affiliation{\MSE}

\author{Malte~R\"osner}
\email{m.roesner@science.ru.nl}
\affiliation{\Radbound}

\date{\today}

\begin{abstract}
    A key advantage of utilizing van der Waals materials as defect-hosting platforms for quantum applications is the controllable proximity of the defect to the surface or the substrate for improved light extraction, enhanced coupling with photonic elements, or more sensitive metrology. However, this aspect results in a significant challenge for defect identification and characterization, as the defect's optoelectronic properties depend on the specifics of the atomic environment. Here we explore the mechanisms by which the environment can influence the properties of carbon impurity centres in hexagonal boron nitride (hBN). We compare the optical and electronic properties of such defects between bulk-like and few-layer films, showing alteration of the zero-phonon line energies, modifications to their phonon sidebands, and enhancements of their inhomogeneous broadenings. To disentangle the various mechanisms responsible for these changes, including the atomic structure, electronic wavefunctions, and dielectric screening environment of the defect center, we combine \textit{ab-initio} calculations based on a density-functional theory with a quantum embedding approach. By studying a variety of carbon-based defects embedded in monolayer and bulk hBN, we demonstrate that the dominant effect of the change in the environment is the screening of the density-density Coulomb interactions within and between the defect orbitals. Our comparative analysis of the experimental and theoretical findings paves the way for improved identification of defects in low-dimensional materials and the development of atomic scale sensors of dielectric environments.
\end{abstract}

\maketitle

\section{Introduction}

  Point defects in semiconductors and insulators have emerged as robust and manipulatable quantum systems for applications such as qubits for quantum computers \cite{Weber2010,Kane1998,Pla2012,Wu2019}, single-photon emitters (SPEs) for quantum communication \cite{Aharonovich2011_RPP,Aharonovich2011}, and nanoprobes for quantum metrology \cite{Schirhagl2014}. In this context, two-dimensional van der Waals (2DvdW) bonded compounds have been proposed as promising host materials. One key benefit arises from the potentially atomically perfect surfaces of these materials. Quantum defects may reside in close proximity to the surface without suffering instabilities in their optoelectronic and/or coherence properties due to surface dangling bonds, adsorbates, or local charge variations. This has several advantages for creating sensors of local fields at the nanoscale, better extraction efficiency for defect SPEs, and direct imaging or manipulating defects by scanning probe techniques \cite{Lee2015, Lee2019}. The most developed of such host material is hexagonal boron nitride (hBN)\cite{Kubota2007,  Cassabois2016}. It is widely available \cite{Wang2017} and its large band gap is conducive to the formation of deep quantum levels. There have been many reports indicating hBN hosts SPEs~\cite{Tran2016, Tran_Acs2016,Koperski_optics2018, Turiansky2019} with some allowing for spin manipulations within the negatively charged boron vacancy defects~\cite{Exarhos2019,Ivady2020, Guo2022}. Specifically, it has been shown experimentally \cite{Koperski2020, Mendelson2021} and theoretically \cite{Huang2012, Maciaszek2022, Wang2021,Huang2022} that carbon impurities in hBN can give rise to a plethora of defect centres with a wide variety of properties attractive for quantum applications.

  The existence of defects in proximity to surfaces and interfaces also presents a significant challenge. Quantum technologies often require arrays of defects with identical properties and though the surfaces and interfaces of the substrate can be atomically clean, proximity to them may change the properties of the defect, including emission energies and linewidths \cite{Jungwirth2016}. For example, hBN films are usually multilayer and range from tens of nanometers to a few atomic layers. Even a defect with the same chemical composition may exhibit different experimental signatures depending on its position with respect to the interface, surface, or edge of the film. A quantitative understanding of these effects is necessary for defect identification and quantum applications.

  In order to understand the influence of surfaces and interfaces on the properties of defects in hBN and other 2DvdW materials, several effects must be disentangled. First, the different distances to surrounding atoms or vacant spaces may change the structure of the defect, especially if it involves displacements of atoms out of the 2D plane. Such structural modifications will also impact the electron-lattice coupling responsible for the phonon broadening of optical transitions. The electronic wavefunctions associated with the defect states may also evolve according to the variations in symmetry and environment. Additionally, the alteration of the dielectric screening environment experienced by the defect will modify the Coulomb interactions between electrons occupying the defect levels. The latter has been recognized before as an important contribution to correlation effects in layered materials, such as for exciton formation \cite{raja_coulomb_2017, Andersen2015, steinhoff_exciton_2017} and magnetism \cite{soriano_environmental_2021}.
  The relative importance of all of these factors is expected to differ between defects with different atomic and electronic structures of the ground and excited states.

  In this work, we use a combination of experimental characterization, via optical and scanning probe spectroscopy, and first-principle theoretical analysis to elucidate the influence of the environment of the defects in 2DvdW materials on their properties. We inspect the environmental sensitivity of the carbon impurity and native defect centres in hBN, as intentional carbon doping allows us to create reproducible defect centres with stable photoluminescence (PL) resonances. Thus, we can control the defect density and the proximity of the defects to the surface by isolating films of varied thicknesses. We demonstrate the evolution of the optical spectra from ensembles of specific defect types in $\sim50$~nm thick hBN films to individual defects in $\sim10$ atomic layers of hBN. We interpret the modification of the PL spectra based on a combination of density-functional theory (DFT) calculations and a recently-developed quantum embedding approach~\cite{Muechler2021}, supported further by scanning tunneling spectra (STS). We apply the theoretical approaches to defects in monolayer and bulk hBN, which allows us to disentangle the various effects on defect properties arising from the surrounding environment.
  
  We find that the dominant effect of the environment is the change in dielectric screening, which affects the intra- and interorbital density-density Coulomb interaction between electrons in defect states. The enhanced screening in bulk form of the crystal significantly reduces the energy of intradefect optical transitions if they depend on such interactions. In addition, we propose that this sensitivity to the dielectric environment can be utilized as a sensor for local dielectric constants when brought into contact with surfaces, adsorbates, or liquids. We highlight that the optical transitions of the defects based on the substitutional carbon dimer have the potential for such applications.
  
  The paper is organized as follows. In Sec.~\ref{sec:exp} we describe our experimental optical and scanning-probe measurements, which motivate the systematic theoretical study of environmental effects on defects given in Sec.~\ref{sec:theory}. In Sec.~\ref{sec:disc} we discuss the main results and the implications for defect identification/characterization, as well as the use of defects to detect changes in a dielectric environment.

\section{Experimental Characterization of Carbon-enriched \lowercase{h}BN \label{sec:exp}}

  In this study, we characterise ultra-pure hBN crystals grown via high-pressure-temperature gradient methods. Carbon doping is achieved post-growth by annealing the crystals in a graphite furnace at 2000$^{\circ}$C for one hour, which gives rise to multiple optically active defect centres~\cite{Koperski2020}. Details on the growth and sample fabrication are given in the Methods Section.
    
  The overall goal of our experimental characterization of these samples is twofold. Firstly, we aim to determine the general properties of the defects that are present to motivate the choices of defects for our theoretical study. Secondly, we aim to inspect how the environment changes the experimental signatures of these defects.

\subsection{Scanning probe measurements}
\begin{figure}
        \centering
        \includegraphics[width=\columnwidth]{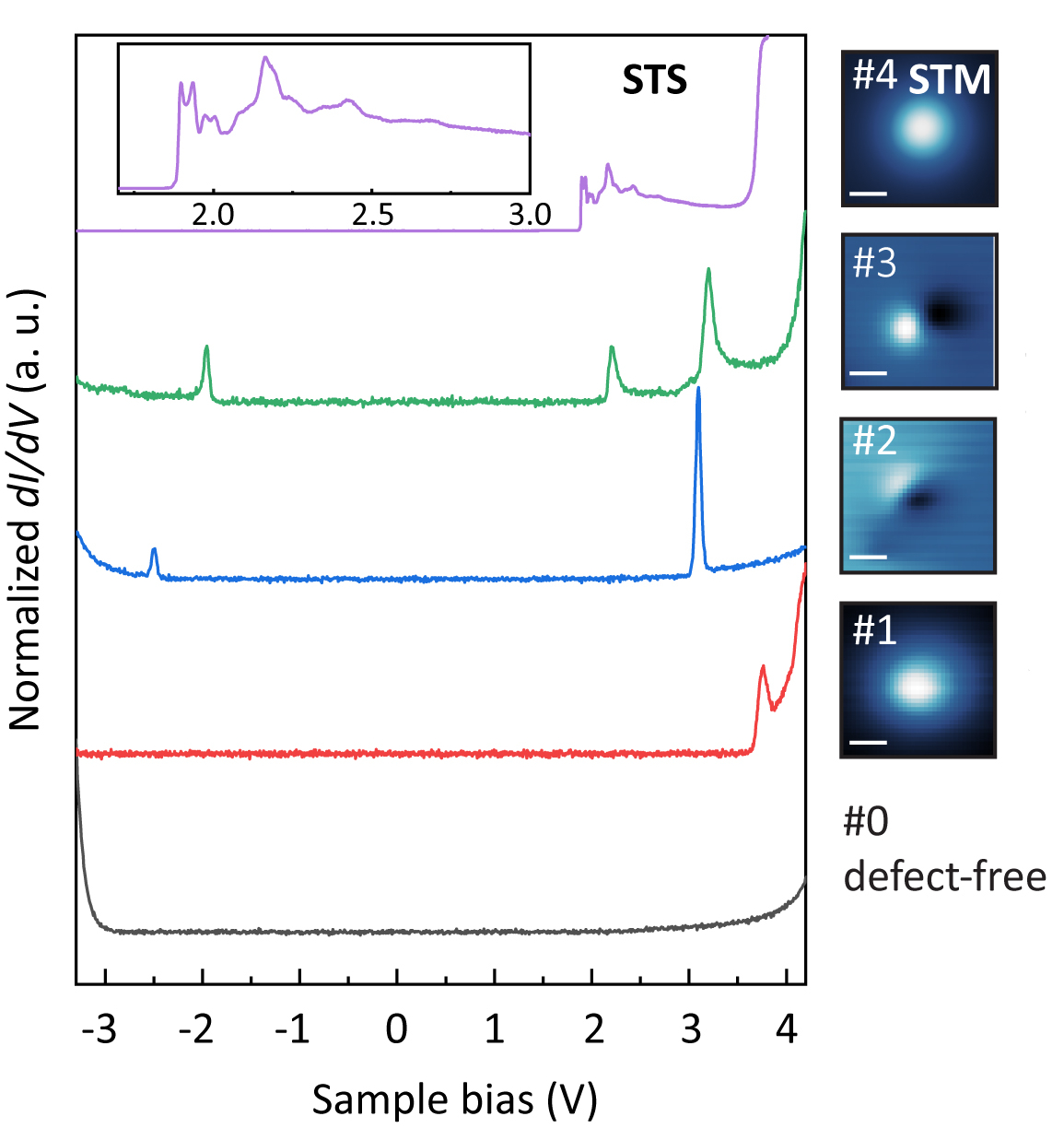}
        \caption{Scanning tunneling spectra (STS) for defects located in 3-layer thick carbon-doped hBN film, which were initially identified through scanning tunnelling microscopy (STM). The STM images are presented next to the corresponding STS curves. All experiments were performed at the temperature $T=77~K$, except for the defect \#4, which was inspected at $T=4.7~K$. STM images of defects \#1$-$\#3 were measured under the conditions $V_s= 5~V$, $I_t= 30~pA$. For defect \#4, STM image was obtained under the conditions $V_s= 4.5~V$, $I_t= 100~pA$. The STS \#0 was measured on a pristine hBN area free of defects. The inset constitutes the magnified image of the STS curve of defect \#4, which demonstrates the fine structure of the vibrionic response arising due to the electron-phonon coupling. The scale bar in the STM images corresponds to 2 nm. \label{fig:sts} }
    \end{figure}

  To address the first objective, we perform scanning-tunneling microscopy/spectroscopy (STM/STS) measurements on a three-layer-thick sample. We can correlate defects observed directly via STM with features of the electronic structure via STS that we expect to see in our theoretical calculations. This allows us to get an overview of the properties of relevant defects in our samples, which is important for the next theoretical steps, as there is a huge variety of carbon-related defects that may occur in hBN \cite{Wang2021,Maciaszek2022,Huang2022, Huang2012}. We present some representative cases in Fig.~\ref{fig:sts}. The STM data (panels on the right) demonstrates well-isolated defect centres that appear as symmetric or asymmetric features in the colour maps. In the $dI/dV$ STS spectra we see that several of these defects (labelled \#1 to \#3), involve resonances near the hBN conduction or valence bands, consistent with simple defects, such as carbon substitutions for boron or nitrogen vacancies~\cite{Attaccalite2011, Auburger2021}.  Accordingly, we start with simple defects in our theoretical modelling discussed in Sec.~\ref{sec:theory}.

\subsection{Photoluminescence spectroscopy \label{sec:PL}}

  We now analyze the optical PL response of carbon-doped hBN in the visible and near-infrared spectral range. We study the effect of the defect environment by considering two samples. The first is bulk-like, characterized by a thickness of about 50~nm. Here, the defects are more likely embedded within the bulk of hBN rather than being localized at the surface or the interface with the SiO$_2$ substrate. The second sample is about 10 hBN layers in thickness, such that on average the defects are in closer vicinity to the surface/interface. In all cases, optical and scanning probe measurements were conducted far from the edges of the samples.

    \begin{figure*}[ht!]
        \centering
        \includegraphics[width=0.99\textwidth]{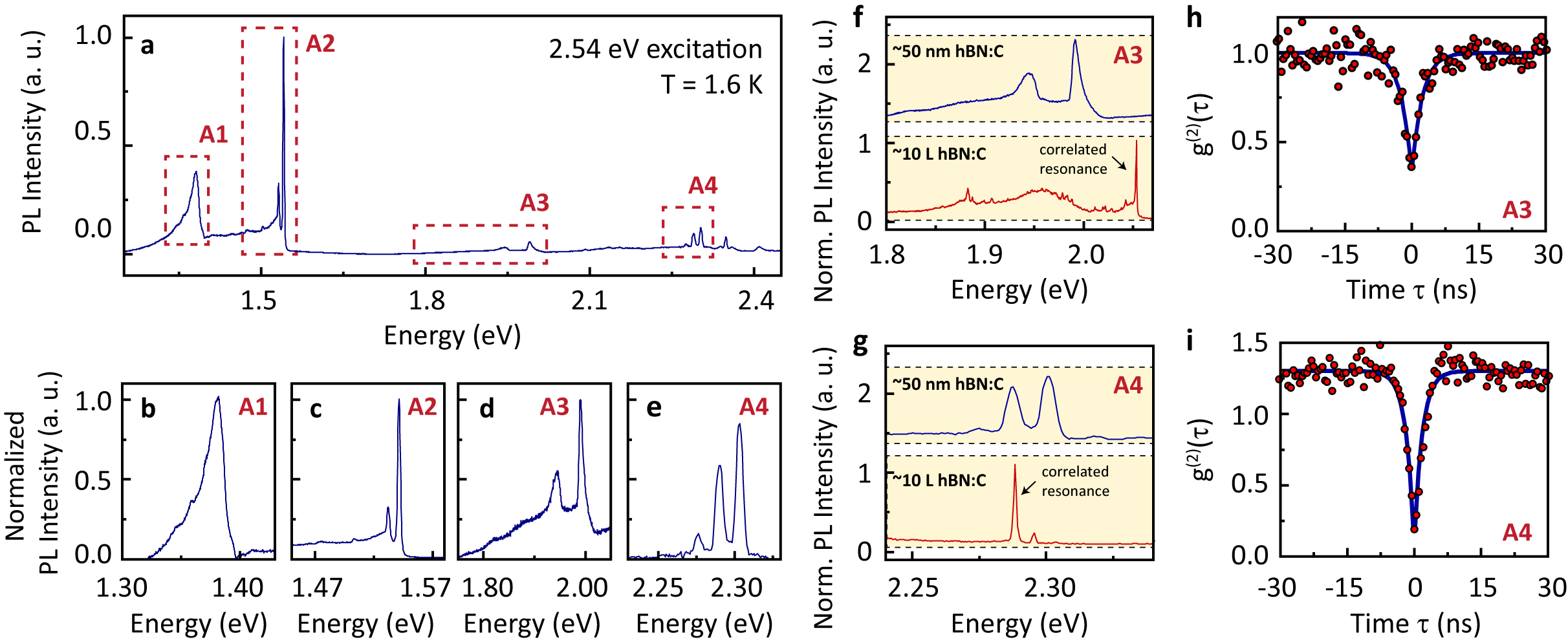}
        \caption{The PL spectra of bulk (about 50~nm thick) hBN:C film measured at low temperature (1.6~K) in microscopic backscattering geometry (a). Four distinct spectral features can be identified, which we attribute to specific defect centres active in near-infrared (A1 and A2) and visible (A3 and A4) spectral regions. The spectra of the specific defects (b-e) are normalized towards uniform maximum intensity for ease of comparison. A comparative representation of PL spectra of hBN:C in the bulk limit ($\sim$50~nm thick) and a few layers' limit (about 10 layers) is presented for defects A3 (f) and A4 (g). The second-order photon correlations $g^{(2)}(\tau)$ measurements done for the emission resonances in 10-layer thick films demonstrate an antibunching indicative of single photon emission for defects A3 (h) and A4 (i).\label{fig:PL}}
    \end{figure*}

  We first focus on defect centres in the bulk-like hBN:C films. In Fig.~\ref{fig:PL}(a) we show the PL spectrum together with a comparative analysis of the four most pronounced defect centres given in Fig.~\ref{fig:PL}(b-e). Defect A1 displays emission in form of a broad asymmetric band indicative of strong electron-phonon coupling. The emission spectra of defects A2 and A4 are dominated by narrow resonances characteristic of centres exhibiting atomic character \cite{Koperski2020}. Defect A3 displays a spectrum typically observed for more complex defect structures such as nitrogen-vacancy centres in diamond, where the well-pronounced dominant zero-phonon line (ZPL) is accompanied by lower-energy phonon sidebands \cite{Koperski2020}. 
  
  We can characterize the luminescence signatures of the bulk hBN defect centres in the framework of a simple one-dimensional Franck-Condon model~\cite{Stoneham}. In that model, the lineshape is governed by the coupling of the electronic transition to a single effective vibrational mode and is parameterized by the Huang-Rhys factor \shr{}, which gives the average number of phonons emitted in an optical transition. \shr{} depends on the difference in atomic structure between the electronic states involved in the transition, and the curvature of the potential energy surface for each state with respect to displacements of the effective vibrational mode. Experimentally, \shr{} can be determined from the ratio of the spectrally integrated intensity of the ZPL and the intensity of the total defect emission (i.e., the Debeye-Waller factor $w = I_\text{ZPL}/I_\text{total}$) using the phenomenological relation $S_{\text{HR}} = -\ln(w)$. In Table~\ref{tab:1} we present $w$ and \shr{} for the defects in bulk-like hBN. We see that they are characterized by small to medium strength of the electron-phonon coupling. For A1, the value of \shr{} should be taken as a lower bound since the ZPL is not discernible. Similarly, the phonon sideband for A4 is not distinctive, and thus we report \shr{} near zero.  
  
    \begin{table}
        \caption{Experimentally measured Debye-Waller and Huang-Rhys factors of defects A1-A4 responsible for the luminescence peaks in Fig.~\ref{fig:PL}(a).\label{tab:1}}
        \begin{ruledtabular}
            \begin{tabular}{cccc} 
                 & ZPL & Debye-Waller factor & Huang-Rhys factor \\
                \hline
                A1 & 1.38 eV & $\leq 0.032 \pm 0.006$ & $\geq 3.4 \pm 0.2$ \\
                A2 & 1.54 eV & $0.29 \pm 0.08$   & $1.2 \pm 0.3$ \\
                A3  & 2.00 eV & $0.12 \pm 0.03$  & $2.1 \pm 0.2$ \\
                A4 & 2.31 eV & $1.0$          & $\sim0.0$ (atomic)
            \end{tabular}
        \end{ruledtabular}
    \end{table} 
    
  Since these defect centres yield reproducible and well-recognizable spectral characteristics, we can directly compare defects in the bulk-like and few-layer samples. In Fig.~\ref{fig:PL}(f,g) we show the corresponding PL spectra for energies around A3 and A4 in bulk and 10-layer thick hBN:C films. Importantly, we find for both hBN thicknesses PL resonances at similar but not identical emission energies. In the A3 energy range, we see that the ZPL blueshifts by about 50~meV from the bulk to the 10-layer sample, which is accompanied by a reduction of the ZPL linewidth (reduction of $I_\text{ZPL}$) indicative of the removal of inhomogeneous broadening and by an increasing Huang-Rhys factor \shr{} from 2.1 to 3.2. As the phonon sideband is not drastically affected, we attribute the enhanced \shr{} to the reduced inhomogeneous ZPL broadening.
 
  The two-peak structure of A4 is visible for both hBN thicknesses and shows a linewidth reduction, without clear signs of a blueshift and the same value of \shr{}  $\simeq 0$. For both defect types, in the few-layer sample, we confirm that we are probing individual defects~\cite{koperski2021,Koperski_optics2018,tran2016robust} by measuring the antibunching in the second order correlation function $g^{(2)}(\tau)$ [see Fig.~\ref{fig:PL}(h,i)]. 
 
  These optical measurements explicitly reveal the significant changes in defect properties, but the particulars vary from defect to defect. This observation provides the motivation for our theoretical investigation of environmental effects on different defects in the next section. Since the attribution of such optical signals with specific defects is still controversial, we will discuss general defects that are likely to exist in hBN:C samples \cite{Wang2021,Maciaszek2022,Huang2022, Huang2012}, and provide further commentary on how our results could be used to aid defect identification in Sec.~\ref{sec:disc}.

\section{Theoretical modelling \label{sec:theory}}

    \begin{figure}[ht!]
        \centering
        \includegraphics[width=0.99\columnwidth]{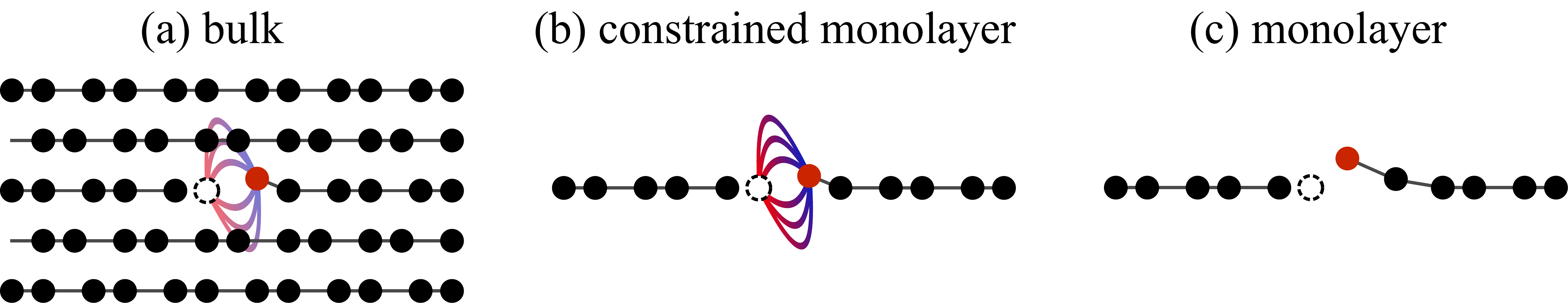}
        \caption{Schematic of theoretical situations used to isolate different environmental effects to impurity complexes (vacancies and C substitutions, the latter in red) embedded in 2DvdW hBN hosts (black). (a) Bulk: defect in the infinite bulk environment. (b) Constrained monolayer: defect in a monolayer constrained to the monolayer structure as in the bulk case. (c) Monolayer: defect with fully relaxed structure in monolayer. In (a) and (b) we sketch intradefect electric field lines to highlight the difference in the screening environment. Note in (c) that the impurity positions are different.
        \label{fig:theory_model}
        }
    \end{figure}

  We proceed by analyzing how the defect structure and the dielectric environment of the hBN host material control intradefect excitations and the Huang-Rhys factors of these excitations. We consider a variety of charge-neutral native and C-containing defects in three different structures: bulk ($P6_3/mmc$), constrained monolayer, and free-standing monolayer hBN, as depicted in Fig.~\ref{fig:theory_model}. The bulk and monolayer structures are fully relaxed, while the constrained monolayer structure corresponds to the impurity embedded in a monolayer but with the same structure as in bulk. With these three structures, we are able to theoretically disentangle the effects of screening and relaxations going from bulk to monolayer hBN. We note that well-controlled impurities in free-standing monolayer hBN are difficult to realize in realistic samples, nevertheless, it serves as a theoretical limit of the strongest environmental sensitivity of defect centres.
  We combine first-principles DFT calculations with the quantum embedding scheme described in Ref.~\onlinecite{Muechler2021}, which allows us to systematically study how changes in the environment and modifications in the impurity structure in the ground and excited states affect the impurity properties. 

\subsection{Computational approach}

  The effect of the environment on the atomic relaxations around the defect is treated at the DFT level and constrained DFT calculations are used to relax the defect structure in excited electronic states. We furthermore estimate how the electron-phonon coupling associated with defect transitions is affected by evaluating the Huang-Rhys factor \shr{}, introduced in Sec.~\ref{sec:PL}. A full picture of the vibrational coupling requires determining $S_{\text{HR}}$ for all relevant modes in the system, which is a challenging task for transitions with moderate $S_{\text{HR}}\sim 3$ as the defects observed in PL \cite{Alkauskas2014,Alkauskas2021}; however, often a qualitative picture of the strength of the electron-phonon coupling can be obtained by considering a single effective mode corresponding to the structural difference between the ground and excited electronic states \cite{Alkauskas2016_Tut,Alkauskas2012}. Under this assumption, and presuming the vibrational coupling is equal in the excited and ground state,  $S_{\text{HR}}= E_{\text{FC}}/\hbar \omega$, where $E_{\text{FC}}=(E_{\text{abs}}-E_{\text{em}})/2$ is the Frank-Condon relaxation energy (half of the difference between vertical absorption and emission energies), and $\omega$ is the frequency of the mode \cite{Stoneham,Alkauskas2016,Alkauskas2016_Tut}. In addition, if we assume that the mode is harmonic, we can write $S_{HR} =  \frac{\Delta Q}{\hbar} \sqrt{\frac{E_{\text{FC}}}{2}}$, where $\Delta Q$ is the absolute change of the nuclear coordinate between the ground and excited defect state indicative of the renormalization of the inter-atomic bond strength.

  The electronic effects of the environment will be elucidated using an embedding approach within which we map the complicated many-electron problem of the defect in hBN to an effective Hamiltonian with only a small number of defect-related states $i,j,k,l$ of the form:
    \begin{equation}
    \begin{split}
    \label{eq:Hubbard_model}
        H&=-\sum_{ij,\sigma}(t_{ij}c^\dagger_{i\sigma}c_{j\sigma}+ \text{H.c.})
        \\
        &+\frac{1}{2}\sum_{ijkl,\sigma\sigma^\prime}U_{ijkl} c^\dagger_{i\sigma}c^\dagger_{j\sigma^\prime}c_{l\sigma^\prime}c_{k\sigma}\\
        &-H_{\text{DC}}-\mu\sum_{i,\sigma} c^\dagger_{i\sigma}c_{i\sigma} ,
        \end{split}{}
    \end{equation}
    where $\mu$ is the chemical potential to control the occupation of the defect states, $\sigma$ is the spin, and $c_i^\dagger$ and $c_i$ are corresponding electronic creation and annihilation operators. $H_\text{DC}$ denotes the so-called double counting correction term~\cite{Muechler2021}. Diagonalizing this Hamiltonian results in many-body energies and wavefunctions of intradefect excitations.

    The ``active space'' of defect-related states is isolated from the host hBN electronic structure via the construction of Wannier functions, which allows us to determine the hopping matrix elements $t_{ij}$. Changes in the hopping matrix elements reflect changes in the properties of the defect states on the DFT level. The screened Coulomb matrix elements $U_{ijkl}$ are computed in the static constrained random-phase approximation (cRPA)~\cite{cRPA} level, which takes the environmental screening into account. Comparing the \emph{unscreened} (bare) Coulomb matrix elements $v_{ijkl}$ (which are only sensitive to changes in defect Wannier functions) with $U_{ijkl}$ gives us insight into how the dielectric environment affects the intradefect transition energies. 
    For computational details, see the Sec.~\ref{sec:methods}.
    
    We can also obtain \efc{} from the intradefect transitions calculated from the many-body energies of Eq.~(\ref{eq:Hubbard_model}). The procedure that makes this possible is to calculate the transition energy in both the ground state geometry and the excited state geometry. Under the assumptions discussed above, half of their difference gives \efc{}.

\subsection{Single carbon impurities}

    \begin{figure}
       \includegraphics[width=1\columnwidth]{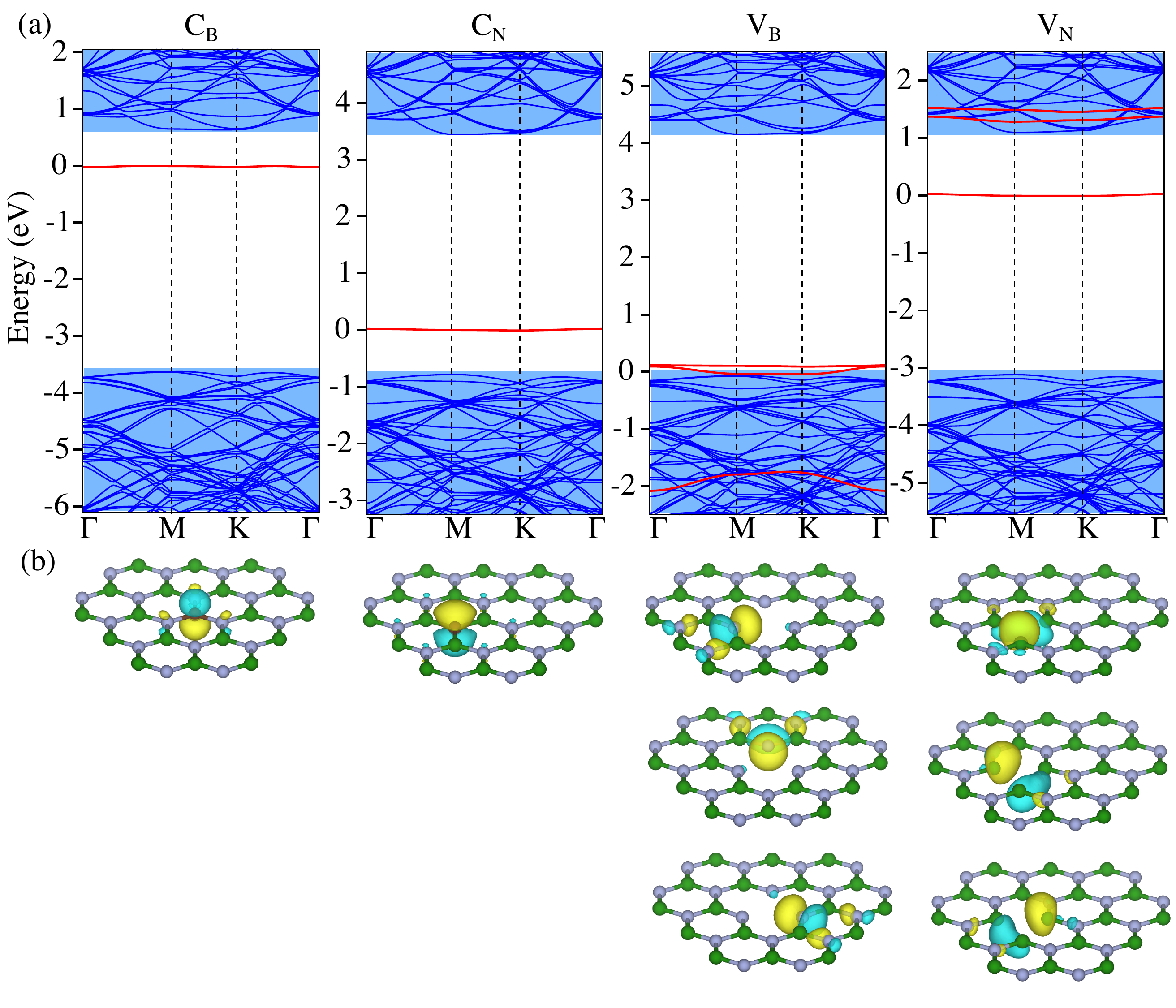}
        \caption{\label{fig:CB_CN} (a) DFT band structure of hBN bulk in the presence of carbon substitution centres and vacancies. Defect states are highlighted with red colours, parametrized using Wannier functions (b).  }  
      \end{figure}
      
    We start our theoretical discussion with single carbon impurities either replacing a boron (\cb{}) or a nitrogen (\cn{}) site, as depicted in Fig.~\ref{fig:CB_CN}. Both yield single occupied in-gap states, close to the hBN conduction band (\cb{}) or valence bands (\cn{}), in good qualitative agreement with the STS/STM data shown in Fig.~\ref{fig:sts}.
    
    With only one electron in a single state, there are no intra-impurity excitations possible. Nevertheless, these most simple defects already allow us to study the impact of the environment on the impurity orbitals. For both of these cases, the change in atomic structure between bulk and monolayer is negligible, with C-N (C-B) bond lengths changing by 0.003~\AA{} (0.004~\AA{}).
    
    We construct localized Wannier orbitals for both C impurities, as depicted in Fig.~\ref{fig:CB_CN}(b). These $p_z$-like Wannier functions allow us to calculate the on-site energies $t_{ii}$, bare ($v=v_{iiii}$) and cRPA screened ($U=U_{iiii}$) Coulomb matrix elements, which we list in Tab.~\ref{tab:Single_impurity}.
    We see that the difference in the bare Coulomb matrix elements between monolayer and bulk is quite small ($\sim 4$\%), indicating that the single-particle electronic structure [i.e., the shape of the $p_z$-like Wannier functions, see Fig.~\ref{fig:CB_CN}(b)] does not change significantly between monolayer and bulk.
    However, $U$ changes significantly, increasing by more than 50\% in the monolayer case compared to bulk. We can cast this into a change in the effective dielectric constants $\varepsilon = v/U$ felt by the electrons in the defect state; for bulk and monolayer hBN we find $\varepsilon \approx 3.8$ and $\approx 2.4$, respectively. Note that this trend follows the out-of-plane dielectric constant decrease from bulk to monolayer hBN \cite{Laturia2018}.
    
    \begin{table}
        \caption{Bare ($v=v_{iiii}$) and screened ($U=U_{iiii}$) Coulomb matrix elements (in eV) of single C impurities. $\varepsilon=v/U$ is the effective dielectric constant. 
        \label{tab:Single_impurity}}
        \begin{ruledtabular}
            \begin{tabular}{lccc|ccc} 
                & \multicolumn{3}{c|}{monolayer} & \multicolumn{3}{c}{bulk} \\
                & $v$ & $U$ & $\varepsilon$ & $v$ & $U$ & $\varepsilon$ \\
                \hline
                \cb{}  & 6.27 & 2.65 & 2.4 & 6.54 & 1.70 & 3.9 \\
                \cn{}  & 7.25 & 2.85 & 2.5 & 7.04 & 1.87 & 3.8 \\
            \end{tabular}
        \end{ruledtabular}
    \end{table} 
    
\subsection{Boron and nitrogen vacancies \label{sec:B_N_vac}}
    
  Boron and nitrogen vacancies, \vb{} and \vn{}, yield more complex impurity states, as illustrated in Fig.~\ref{fig:CB_CN}. The \vn{} defect gives rise to a half‐filled level below the conduction band accompanied by two additional nearly degenerate empty defect levels within the edge of the conduction band, which is also in qualitative agreement with the STS/STM data of impurity $\#1$ in Fig.~\ref{fig:sts}(a). The \vb{} defect creates three levels originating from the dangling bonds due to the missing B atom. Two of these states are in close vicinity to the valence band edge, nearly degenerate, and host together one electron, while the third level is deeply buried within the valence band. In these cases, the hopping matrix $t_{ij}$ has a $3\times3$ form, while the screened Coulomb interaction $U_{ijkl}$ is a rank-4 tensor with three elements per dimension. To compare the Coulomb interactions, we define the average intra-orbital density-density interaction $U_0=N_\text{orb}^{-1}\sum_i U_{iiii}$, inter-orbital density-density interaction $U_1=[N_{\text{orb}}(N_{\text{orb}}-1)]^{-1}\sum_{i\ne j} U_{ijij}$, and exchange interaction $J=[N_{\text{orb}}(N_{\text{orb}}-1)]^{-1}\sum_{i\ne j} U_{ijji}$, where $N_{\text{orb}}$ is the number of Wannier orbitals. Furthermore, we define the effective dielectric constant $\varepsilon = v^{(d)}_l/U^{(d)}_l$ using the leading eigenvalues of the bare and cRPA screened density-density Coulomb matrices~\cite{Rosner2015}.

  In Tab.\ref{tab:Single_vacancies} we compare all these parameters together with the single-particle energy separation $\Delta E = t_{00} - t_{11}$ between Kohn-Sham eigenvalues for monolayer and bulk hBN hosts. We find that single-particle energies associated with these defects [and thus the hoppings in Eq.~(\ref{eq:Hubbard_model})] change by 2\% for \vb{} and 4\% for \vn{}. In contrast, the screened density-density interactions decrease significantly from monolayer to bulk by more than 50\%. Similar to the screening effects for the C substitutional defects, this is driven by the effective dielectric constants decrease from $\sim3$ in the bulk to $\sim 2$ in the monolayer. Notably, the exchange elements $J$ are nearly unaffected by the changes in the environment as a result of their dipolar character. Indeed, macroscopic dielectric environmental screening mostly affects density-density Coulomb interactions in layered materials and does not significantly affect non-density-density elements \cite{Rosner2015,soriano_environmental_2021}.
  Thus, many-body properties which are strongly affected by density-density Coulomb interactions, such as charge excitations, will be in general most prone to changes in the dielectric environment, e.g., resulting from transition from bulk to monolayer hBN, while exchange interaction driven many-body states, such as spin excitations, will be less affected.
    
\begin{table}[h]
    \caption{Coulomb matrix elements (in eV) of single B/N vacancies. See Sec.~\ref{sec:B_N_vac} for the definition of variables.  \label{tab:Single_vacancies}}
    \begin{ruledtabular}
        \begin{tabular}{l|ccccc|ccccc} 
            & \multicolumn{4}{c}{monolayer} & \multicolumn{4}{c}{bulk} \\
            & $U_0$ & $U_1$ & $\varepsilon$ & $J$ & $\Delta E$ & $U_0$ & $U_1$& $\varepsilon$ & $J$ & $\Delta E$   \\
            \hline
            \vb{} & 5.04 & 2.17 & 2.4 & 0.03 & 1.86 & 3.45 & 1.31 & 3.6  & 0.03 & 1.82 \\
            \vn{} & 3.50 & 3.00 & 2.2 & 0.23 & 1.60 & 2.47 & 2.00 & 3.2  & 0.23 & 1.54 \\
        \end{tabular}
    \end{ruledtabular}
\end{table} 

  This behaviour is visible in the many-body energies of \vb{}, as summarized in Tab.\ref{tab:Single_vacancies_eigenstates}. For both monolayer and bulk hBN hosts, the \vb{} ground state corresponds to four-times degenerate quadruplet state $Q_0$, which is approximately given by three electrons fully occupying the lowest and partially occupying the degenerate single-particle states. From the corresponding excitation energies, we find that nearly all excited states decrease in energy by about 500 to 600 meV (with $Q_1$ as the sole exception) in bulk. This demonstrates that these transitions are decisively affected by the density-density Coulomb interactions. 

  For \vn{} we find a similar trend with doublets as the ground state and the following quadruplet excited state. However, all these states are occupied by a single electron, which eliminates any Coulomb contribution. The differences we see in the excited energies stem from the slightly altered single-particle properties as indicated in Tab.~\ref{tab:Single_vacancies}. 

\begin{table}[h]
    \caption{Many-body states of single B/N vacancies. \label{tab:Single_vacancies_eigenstates}}
    \begin{ruledtabular}
        \begin{tabular}{c|c|c|c} 
            & spin & $E_n$ monolayer (in eV)  &    $E_n$ bulk (in eV) \\ 
            \hline
            \multirow{5}{*}{\vb{}}  &  $Q_0$ &  0     &   0       \\
                                    &  $Q_1$ &  0.51  &   0.63    \\
                                    &  $Q_2$ &  2.82  &   2.31    \\
                                    &  $D_1$ &  3.18  &   2.61    \\
                                    &  $D_2$ &  3.63  &   3.00    \\
                                    &  $Q_3$ &  4.63  &   4.04    \\

            \hline
            \multirow{2}{*}{\vn{}}  &  $D_0$ &  0     &    0       \\
                                    &  $Q_1$ &  1.48  &    1.37    \\
         \end{tabular}
    \end{ruledtabular}
\end{table} 

\subsection{Impurity complexes}

  The picture that emerges from the simple defects discussed in the previous two chapters outlines the key role of differential dielectric screening, which influences density-density Coulomb interactions within and between defect orbitals, inducing a change in the properties upon varying the defect environment. 

\begin{figure*}
   \includegraphics[width=1.05\textwidth]{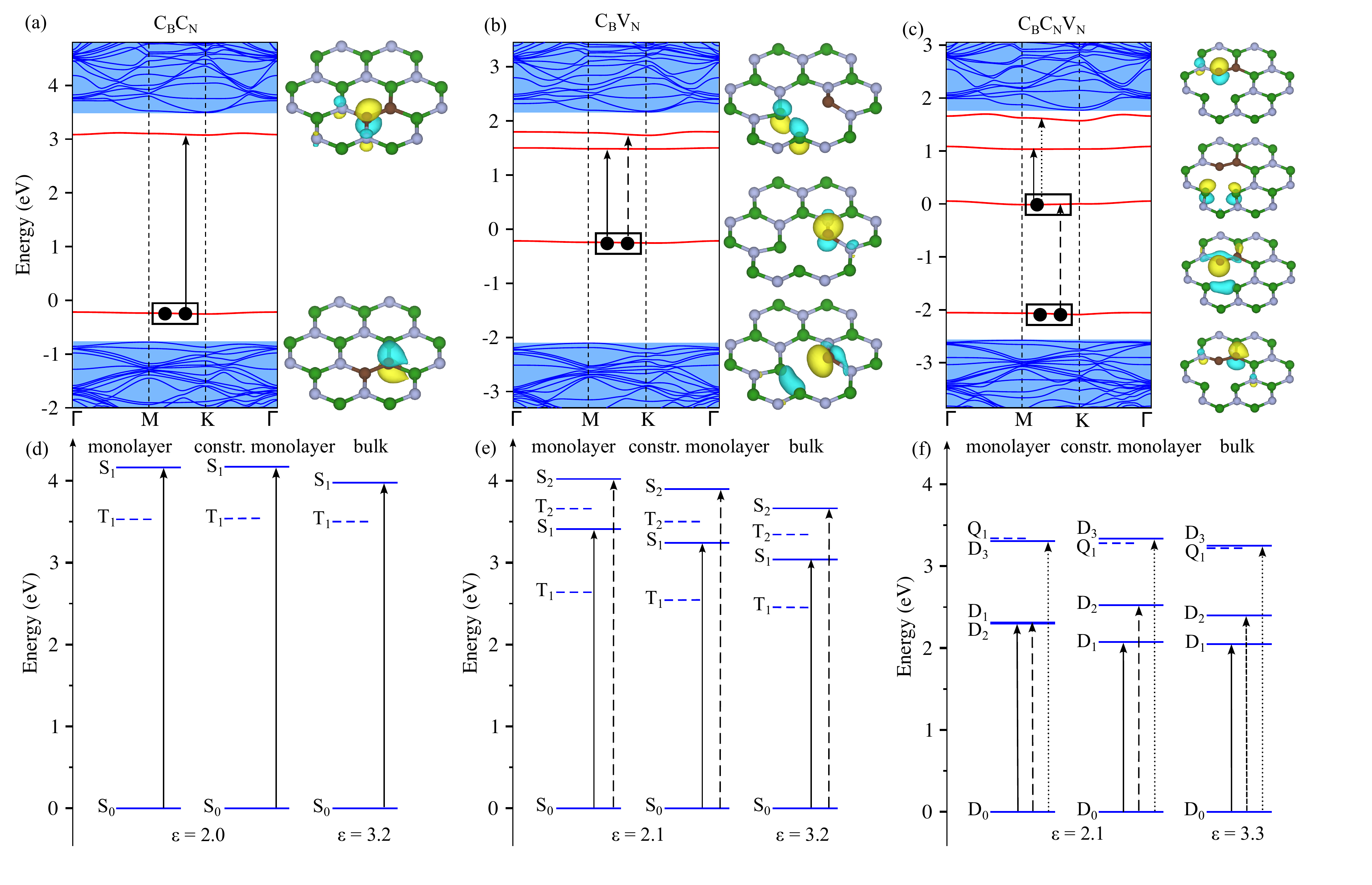}
    \caption{\label{fig:theory} (a-c) DFT band structures of each considered impurity complexes embedded into bulk hBN with highlighted in red impurity states, which were parametrized using Wannier functions.  Black dots demonstrate initial occupations and their modification upon optical excitations. (d-f) Many-body impurity states embedded into bulk, constrained monolayer, and \textit{free-standing} monolayer of hBN. Arrows demonstrate the possible excitations, which properties are given in Table~\ref{tab:monolayer_bulk_evolution}.}
\end{figure*}

  We now consider more complex defect/impurity structures, which have in-gap intradefect transitions with possible technological relevance, and that could be detected by optical means, as exemplified in Sec.~\ref{sec:PL}. 
  Specifically, we consider a carbon dimer (\cb{}\cn{}) replacing nearest-neighbour B and N atoms with carbon, a carbon-vacancy complex (\cb{}\vn{}), and a combination of a dimer with a neighbouring vacancy (\cb{}\cn{}\vn{}). Especially \cb{}\cn{} has recently attracted significant attention, as it was proposed as the origin of the 4.1~eV ZPL single-photon emitter \cite{Era1981,Museur2008,Du2015} observed in hBN based on the energetics of emission \cite{Mackoit2019} and calculations of photoluminescence lineshapes \cite{Linderalv2021,Jara2021}.   All these impurity complexes have modest formation energies~\cite{Huang2022, Maciaszek2022} and are thus likely to occur in both, monolayer and bulk hBN upon C exposure. In Fig.~\ref{fig:theory} we summarize all model details. 

  From the DFT band structure calculations presented in Fig.~\ref{fig:theory} (a-c) we see that all three impurity complexes form in-gap defect states with varying occupations (two electrons in \cb{}\cn{} and \cb{}\vn{} and three electrons in \cb{}\cn{}\vn{}). We use these states as the active space for our minimal modelling [see Supplemental Material (SM) sec. S2 for details]. The resulting lowest excited many-body levels are shown in Fig.~\ref{fig:theory} (d-f) with indications of the ground and excited state types, singlets (S), doublets (D), triplets (T), and quadruplets (Q). We show these energies calculated in the three cases schematically depicted in Fig.~\ref{fig:theory_model}. Specifically, the inclusion of a constrained monolayer, i.e., where the defect structure is kept the same as in bulk but the other atomic layers are removed, allows us to isolate the effects of the dielectric environment from changes in structure from the atomic environment. We thus first comment on the comparison between bulk and constrained monolayer, and then on the effect of relaxations between bulk and monolayer.

  Throughout all impurity complexes, we see a consistent trend of decreasing optical transition energies when changing from the constrained monolayer to the bulk hBN. Upon allowing for further relaxation of the monolayer host system, this trend is mostly preserved. The decreasing excitation energies are more pronounced for singlets than for doublets and are often vanishingly small for triplets and quadruplets,
  which is a direct measure of how much the many-body states are controlled by density-density and exchange Coulomb matrix elements as illustrated for each defect below.

\subsection{Many-Body States}

\subsubsection{C$_B$C$_N$ Many-Body States}

 The \cb{}\cn{} carbon dimer impurity can be approximately modeled as a half-filled Hubbard dimer~\cite{Ocampo2017,Muechler2021}. Analytically, the many-body eigenstates of the Hubbard dimer are solely controlled by the ratio of the local density-density Coulomb repulsion $U$ and the hopping $t$ between the sites. In our case this dimer is formed by localized $p_z$-like orbitals at the two C positions with local $U_0$ and interorbital Coulomb interactions $U_1$, see Fig.~\ref{fig:theory}(a). In this scenario the effective local Coulomb interaction magnitude is approximately given by $U^* = U_0 - U_1$\cite{schuler_optimal_2013,carrascal_hubbard_2015}. Upon increasing the screening from monolayer to bulk, both $U_0$ and $U_1$ are significantly but similarly reduced (e.g., $\Delta U_0 \approx 1.3$~eV and $\Delta U_1 \approx 1.1$~eV), such that the resulting $U^*$ differ by $\Delta U^* \approx 0.2\,$~eV. This corresponds to a relative change of nearly 40\%. As the electronic hopping between the two $p_z$ Wannier functions is affected by only 3\% through the monolayer to bulk transition, we can again state that modifications to the many-body properties are mostly driven by the changes in the dielectric screening. This explains the reduction of the $S_1$ state by about $0.2$~eV. In contrast, the lower triplet states $T_1$ is less affected by this screening, as it depends more strongly on Hund's exchange elements, which are nearly unaffected (see SM Table SIV). Further, the data presented in Fig.~\ref{fig:theory}(d) shows that atomic relaxations of the defect between bulk and monolayer do not play a significant role here (c.f. constrained monolayer and monolayer).

 \begin{table*}[ht]
\caption{The absolute change of configuration coordinates $\Delta Q$ (in $\sqrt{\mathrm{amu}}$ \AA), Franck-Condon energy $E_{FC}$ (in eV) and Huang-Rhys factor \shr{}  for  impurity complexes embedded into free-standing monolayer and bulk hBN. \label{tab:monolayer_bulk_evolution}}
\begin{ruledtabular}
\begin {tabular}{c|ccc|ccc|ccc|ccc|ccc|ccc}
& \multicolumn{6}{c|}{\cb{}\cn{}} &   \multicolumn{6}{c|}{\cb{}\vn{}} &  \multicolumn{6}{c}{\cb{}\cn{}\vn{}}  \\
\hline
& \multicolumn{3}{c|}{monolayer} & \multicolumn{3}{c|}{bulk}  & \multicolumn{3}{c|}{monolayer}  & \multicolumn{3}{c|}{bulk} &  \multicolumn{3}{c|}{monolayer}  & \multicolumn{3}{c}{bulk} \\
\#  & $\Delta Q$  & \efc{} &  \shr{} &  $\Delta Q$  & \efc{}  &  \shr{} & $\Delta Q$ & \efc{} &  \shr{} & $\Delta Q$  & \efc{} &  \shr{} & $\Delta Q$  & \efc{} &  \shr{} & $\Delta Q$  & \efc{} &  \shr{} \\  
\hline
1 &  0.24   &  0.06  &  0.65 &  0.26   &  0.07 &  0.75  &  3.37 &  0.80 &   33.0 &  2.00  &   0.62 &  17.20  &  2.95 &  0.59 & 24.80 & 1.68  &  0.44  & 12.19  \\
2 &         &        &       &         &       &        &  1.51 &  0.23  &   8.0 &  0.78  &    0.18 &  3.67  &  1.79  & 0.65 &  15.78& 1.46  &  0.64 & 12.81   \\
3 &         &        &       &         &       &        &       &       &        &        &         &        &  2.81  & 0.13 &  11.00 & 1.12  &  0.08 &  3.38   \\

\end {tabular}  
\end{ruledtabular}
\end {table*}

\subsubsection{C$_B$V$_N$ Many-Body States}

  For \cb{}\vn{} the spinless DFT Kohn-Sham (KS) ground state is approximately given by two electrons mostly residing in a C-centered $p_x$-like state [see side panel of Fig.~\ref{fig:theory}(b)] with a distortion of the C atom position out of the plane. The fully interacting many-body ground state of this impurity can be well approximated as a single Slater determinant with two electrons of opposite spin in the lowest Kohn-Sham state forming a singlet ground state. The many-body singlet-singlet transitions approximately promote one of these electrons either into a C-centered $p_z$-like state ($S_0 \rightarrow S_1$, which lets the C atom relax back to the hBN plane) or into a delocalized state with two $p_z$-like wavefunctions centered at the neighbouring B atoms ($S_0 \rightarrow S_2$ with reduced out-of-plane C position distortion). The bulk screening reduces the corresponding transition energies by $0.2$~eV and by $0.24$~eV for $S_1$ and $S_2$, respectively [c.f. constrained-monolayer to bulk transitions in Fig.~\ref{fig:theory}(e)]. A similar approximate $U^*$ analysis (using orbital averaged $U_0$ and $U_1$ given in the SM) yields an estimate of $\Delta U^* \approx 0.2$~eV (ca. $20\%$).
  As the single-particle energies only change by less than 5\% and since the Hund's exchange elements are nearly unaffected between the bulk and constrained monolayer hosts, the overall trend in the many-body levels can again be explained based on modifications to the density-density Coulomb matrix elements. Full relaxations towards the free-standing monolayer enhance the out-of-plane C position distortion, which mainly affects the single-particle energies, while screened density-density and Hund's exchange elements are nearly unaffected (see Table SV). The single-particle KS energies change by about $0.1$ to $0.2\,$~eV in the freestanding monolayer with respect to the constrained monolayer, which is the same order of magnitude as the many-body excitation energies change between the two systems.

\subsubsection{C$_B$C$_N$V$_N$ Many-Body States}

This defect complex can be interpreted as a combination of \cb{}\cn{} and \cb{}\vn{}. The KS ground state corresponds approximately to a fully occupied $p_z$-like state centered on the C atom farthest from the vacancy and a half occupied $p_x$-like state centered at the C adjacent to the vacancy. This impurity thus hosts three electrons. Due to the spin degree of freedom of the unpaired electron, the many-body ground state is two-fold degenerate forming a doublet state, whose wave functions can again be well approximated by a single Slater determinant in the band basis (see Table SX). The ground state structure shows an out-of-plane distortion, mostly in the C position. Doublet-doublet excitations approximately promote the unpaired electron from the $p_x$ orbital to delocalized $p_z$-like orbitals at the neighbouring B $(D_0 \rightarrow D_1$ letting the C atom relax back to the plane), one of the electrons from the C $p_z$ to the C $p_x$ $(D_0 \rightarrow D_2$ with enhanced out-of-plane distortions) and the unpaired electron from the C $p_x$ to C $p_z$  $(D_0 \rightarrow D_3$ also letting the C position relaxing back to the plane). Here, the many-body energies are less influenced by the dielectric environment. $D_1$, $D_2$, and $D_3$ change by about $-0.03$, $-0.13$, and $-0.08\,$~eV respectively, although, as before the effective density-density Coulomb matrix element is reduced by about $\Delta U^* \approx 0.15\,$~eV, while the single particle energies are affected by less than 5\% when changing from the constrained-monolayer to the bulk hBN host. Thus only the D$_2$ transition approximately follows $\Delta U^*$, while the other excitations are less affected. We attribute this different behaviour to the difference in the nature of the excitations: $D_1$ and $D_3$ are single electron excitations, while $D_2$ takes place among three electrons and involves significant modifications to the impurity charge density. Under further relaxation towards the free-standing monolayer hBN, which enhances the out-of-plane distortion, the main difference is that the many-body states $D_1$ and $D_2$ become nearly degenerate. This is driven by modifications to the single-particle Kohn-Sham energies, while the screened Coulomb interactions are approximately the same (see monolayer to constrained-monolayer transition in Table SV).

\subsubsection{Huang-Rhys factors}

  We now turn our focus to the Huang-Rhys factors \shr{}. In Tab.~\ref{tab:monolayer_bulk_evolution} we summarize all $\Delta Q$, \efc{} and \shr{} for both the free-standing monolayer and bulk hBN hosts. For both cases, we find that \shr{} corresponding to the excited states in \cb{}\cn{} are considerably smaller than in \cb{}\vn{} and \cb{}\cn{}\vn{}. Quantitatively this is driven by the decisive out-of-plane distortion of the \cb{}\vn{}/\cb{}\cn{}\vn{} ground state which is reduced in their excited states, yielding large $\Delta Q$. This is related to the decisively modified impurity charge densities in \cb{}\vn{}/\cb{}\cn{}\vn{} upon excitation. In the ground states we find an in-plane charge density with $p_x$ symmetry localized partially on carbon and partially on the nearest boron sites in presence of the N vacancy. Hartree terms (i.e. bare Coulomb repulsions) between these two partial densities destabilize the planar crystal structure, shifting the carbon up in $z$-direction in the ground state. Further full relaxation towards the free-standing monolayer limit enhances this out-of-plane distortion, which consequently increases \shr{} via enhancing $\Delta Q$. However, we note that in actual experiments these additional distortions are likely suppressed by the presence of a substrate or capping layer.
  In addition to $\Delta Q$, there is also a slight but consistent trend of \efc{} being less in bulk than in monolayer (\cb{}\cn{} is again the exception, though the change is negligible).

\section{Discussion \label{sec:disc}}

\subsection{Mechanisms for environmental effects on defect properties}

  In Sec.~\ref{sec:exp}, we experimentally showed via PL measurements that the environment of carbon-based defects affected their properties, such as the energy of optical resonances and Huang-Rhys factors \shr{}. In Sec.~\ref{sec:theory}, we explored theoretically the possible mechanisms for these changes by comparing monolayer and bulk hBN hosts. In the following we summarize our findings, focusing mostly on defect complexes with technologically relevant intradefect transitions.

  Firstly, we consider differences in defect structure between monolayer and bulk hBN. These changes are relatively minor when the distortions related to the defect are purely in the hBN plane, as is the case for the single vacancies, single C impurities, and \cb{}\cn{} (see, e.g., Table~\ref{tab:monolayer_bulk_evolution} and SM). For defects with ground-state out-of-plane corrugation (e.g., \cb{}\vn{}), the displacements are larger in monolayer than in bulk hBN hosts as there are no other hBN layers to constrain them (see SM). The effect of the increased distortions on the many-body energy differences and Frank-Condon relaxation energies is relatively modest. However, \shr{} can be significantly larger in freestanding monolayer hBN hosts than in bulk ones for excitations accompanied by out-of-plane to in-plane geometry relaxations, which significantly increase $\Delta Q$ for the corresponding transitions.

  We furthermore show, that changing the hBN host environment mostly affects screened density-density Coulomb matrix elements within the impurity states, which can be reduced in the bulk by up to $1.3\,$~eV, while single particle energies are modified by only up to $0.2\,$~eV, and Hund's exchange Coulomb matrix elements barely change.
  Both the interorbital and intraorbital density-density interactions are reduced significantly in bulk hBN hosts by $\sim 50$\% compared to the monolayer case. Quantifying this change with an effective dielectric constant give $\varepsilon_{\text{bulk}} \approx 3$ to $4$, while $\varepsilon_{\text{ML}} \approx 2$ to $2.5$ depending on the defect. However, the extent to which these changes affect the intradefect many-body energies depends on how much the many-body levels depend on the interplay and compensation of the (partially various) density-density interactions and Hund's exchange parameter $J$. Overall, the resulting trend is that the energy separation of intradefect levels decreases in bulk compared to monolayer, which is, in general, most prominent in charge excitations.

\subsection{Implications for defect identification}

  Our findings have significant implications for the identification of defects in hBN and 2DvdW layered hosts in general. Firstly, we point out that renormalization of optical transitions from interactions and the correct treatment of spin symmetry is important for accurate comparison between experiment and theory, as is clear from comparing the KS single-particle states in Fig.~\ref{fig:theory}(a)-(c) with the many-body states (d)-(e). Here, this was achieved by our embedding approach, which can often also be achieved, e.g., via $\Delta$SCF calculations utilizing hybrid functionals and possibly corrections for spin contamination \cite{vonBarth1979}. A representative example of this observation is \cb{}\vn{}, wherein the single particle picture two transitions of about 2~eV are expected, while in the many-body picture the singlet-singlet transitions appear at energies above 3~eV, and triplet-triplet transition occur at about 1~eV [see Fig.~\ref{fig:theory}(b,e)].

\begin{figure}
      \includegraphics[width=1\columnwidth]{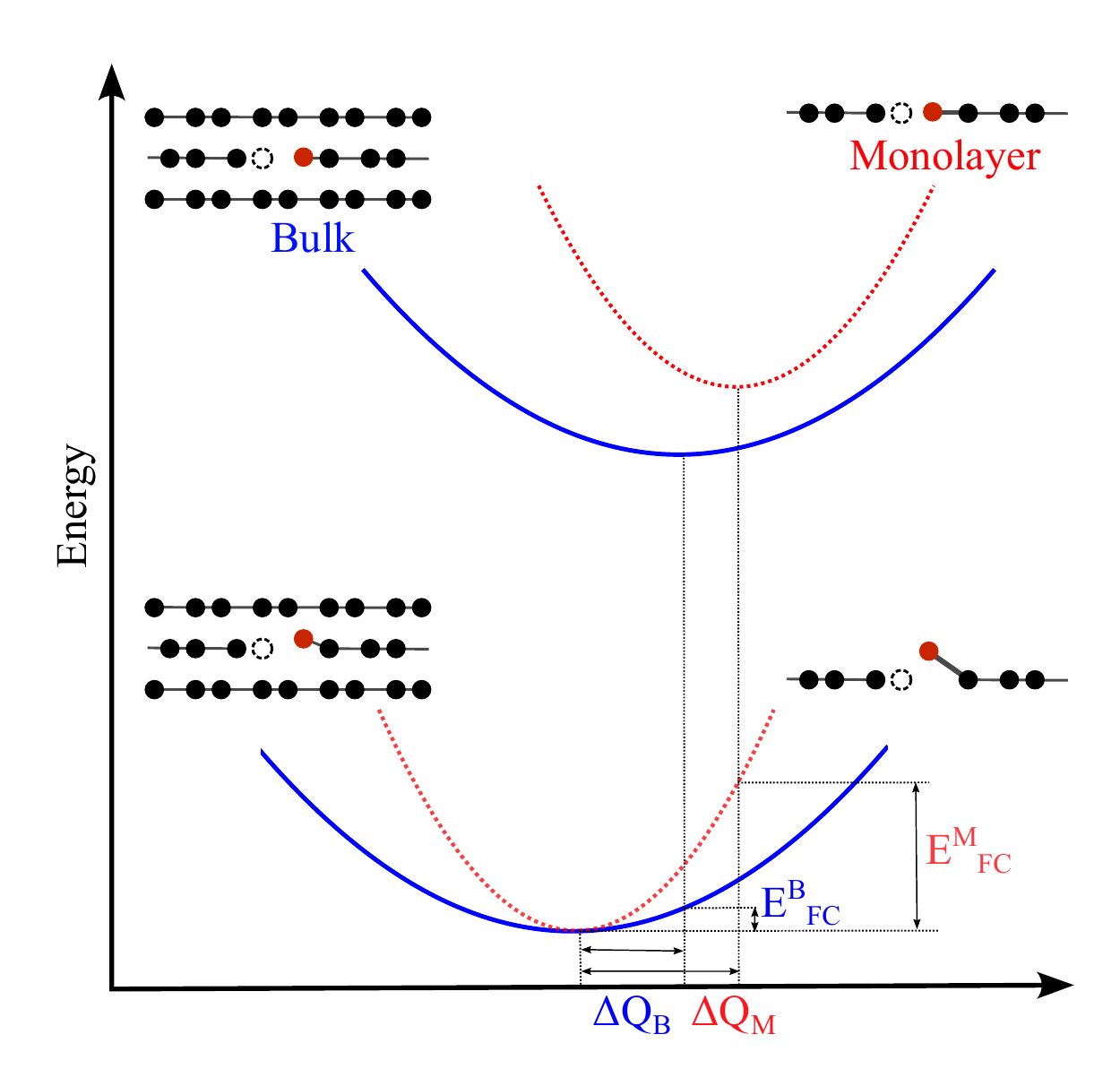}
       \caption{\label{fig:ML_vs_BL} Configuration coordinate diagram for defect complexes embedded to  bulk and  monolayer sample. }
  \end{figure}

  Moreover, we have shown that intradefect transitions within 2DvdW hosts are controlled by additional degrees of freedom, i.e., the surrounding environment of the defects. In our optical measurements this was tuned by the host-material thickness, while in theoretical calculations we considered defects in bulk and monolayer hBN. The general trend of increasing intradefect energies and increasing \shr{} in monolayer compared to bulk hosts are summarized in Fig.~\ref{fig:ML_vs_BL}: the host material can change both, $\Delta Q$ and $E_{FC}$. 
  Thus, for experimental defect identifications within 2DvdW hosts, shifts in many-body energies and changes in \shr{} should be taken into account across samples. On the other hand, such trends can be used to aid defect identification and characterization if a signal can be correlated between samples of different thicknesses, or on different substrates. 

  In light of these results, we can reexamine the defects observed in the PL measurements described in Sec.~\ref{sec:PL}.  In the considered energy range, we observed four types of defects with varying emission energies and Huang-Rhys factors \shr{}, which behave differently upon variation of the hBN film thickness. The defects $A_2$ and $A_4$ have PL resonances in the range of the doublet-doublet transitions related to $V_N$ and $V_B$ containing defects, respectively. However, \shr{} for these defects is much larger than observed experimentally \cite{Li2022}. $A_3$ likely involves a defect similar to the carbon dimer, which might be coupled to neighbouring vacancies, as it shows a relatively small \shr{} and strong sensitivity of the PL resonance energy to the hBN film thickness. Defect $A_1$ is characterised by the largest \shr{} and lowest emission energy (within our series). Assuming that $A_1$ is a simple defect, we could attribute it to the triplet-triplet transition within \cb{}\vn{}. However, it is also plausible that this defect does not belong to the defect space considered here. Interestingly, the investigated defects exhibit vanishing small Zeeman splitting, when the magnetic field is applied perpendicularly to the hBN plane (see SM, Fig. S1). This observation may indicate that the optically active transitions are dominated by singlet-singlet transitions, however, the transitions between the parallel branches of the spin-split doublets and triples cannot be excluded.

  Finally, we point out that we have not explicitly considered the effect of a substrate in this work. Depending on the substrate the impact on the defect properties may change significantly. On one hand, we would expect that the substrate could confine out-of-plane lattice distortions caused by the defect similar to the additional layer within bulk hBN hosts. 
  On the other hand, the dielectric screening from certain substrates might be significantly larger compared to the moderate enhancement from monolayer to bulk hosts. Substrates with significant lattice contributions to the dielectric susceptibility may result in effective relative dielectric constants of 20 and above~\cite{PhysRevB.19.3593,robertson_high_2004}.

\subsection{Defects for detecting local dielectric susceptibility}

  We can use the sensitivity of defects in 2DvdW materials as probes of the local dielectric environment of, e.g., other surfaces, adsorbed molecules, or confined liquids \cite{Fumagalli2018}. An ideal defect for such applications should exhibit an optically accessible intradefect transition that has a strong dependence on the dielectric environment. Also, the defect should be characterized by a spectrally narrow optical signal providing good sensitivity and resolution for determining the properties of the dielectric environment. Thus a small \shr{} is desired. Finally, it would be ideal if the defect properties were relatively insensitive to other environmental stimuli such as external magnetic or electric fields.

  Experimentally, defect $A_3$ appears as the best candidate for detecting the local dielectric constant. The PL emission resonance shifts by $\Delta E = 60.1$~meV between the 50~nm thick hBN film and 10-layer-thick hBN film. We can estimate the upper bound of the modification of the relative dielectric constant sensed by the defect within the two films based on our calculation to be $\Delta \varepsilon < 2$. Therefore, we can conclude that our sensitivity towards the local dielectric constant is better than $\Delta \varepsilon / \Delta E < 0.03 $~meV$^{-1}$. The resolution is limited by the linewidth of the PL resonance $\delta$, which also depends on the layer thickness. In the bulk form, the line exhibits its largest broadening, $\delta = 7.4$~meV, which corresponds to the change of the relative dielectric constant $\Delta \varepsilon  < 0.22$. In the monolayer form, $\delta = 1.5$~meV, which yields $\Delta \varepsilon  < 0.05$. 

  From our theoretical point of view, the singlet-singlet transition in \cb{}\cn{} is a promising candidate. Although it has rather high energy, its ZPL has been detected and identified previously \cite{Era1981,Museur2008,Du2015, Mackoit2019}. Here, we have demonstrated that it is highly sensitive to the defect environment. Also, it has a Huang-Rhys factor that does not change with the dielectric environment, so its ZPL will remain sharp. Finally, it is a singlet-singlet transition, such that it likely will not be affected by stray magnetic fields. It may however shift under the application of electric fields since the defect has a relatively low symmetry (point group $C_{2v}$). 

  Therefore, carbon defect centres in hBN can be used as sensitive detectors of local dielectric constants, addressing the need of measuring the environmental effects on a nanoscale. 

\section{Conclusions}

Using a combination of experiment and theory, we have elucidated the effect of the environment on defect properties in hBN. We show via PL measurements of few-layer and bulk-like samples, that carbon-based defects exhibit shifts in ZPL energies, as well as changes in phonon sidebands and ZPL lineshapes, which can be quantified via Huang-Rhys factors. Using first-principles theory and embedding methods applied to monolayer and bulk hBN, we show that the key effect of the environment arises from modifications in the effective local dielectric screening acting on the correlated impurities. This alters the inter- and intra-orbital impurity Coulomb interactions and plays a role in reducing optical emission energies for intradefect transitions in bulk versus mono/few-layer hBN hosts. These effects  must be taken into account when performing defect identification via a comparison between experiment and theory. A plausible application of our findings is a detector for local dielectric constants, easily integrable with solid, soft, and liquid matter systems in pristine and/or functionalized form.

\section*{Methods \label{sec:methods}}

  \textbf{Crystal growth} The pristine ultra-pure hBN crystals have been grown via the high-pressure temperature-gradient method. A part of the crystals from a single growth batch was annealed in a graphite furnace at a temperature of 2000$^{\circ}$C for an hour.

  \textbf{Sample fabrication} The hBN:C films were isolated through mechanical exfoliation of bulk crystals onto 300 nm thick Si/SiO$_2$ substrates. In the transfer process, the substrates were heated to 50$^{\circ}$C to increase the yield of thin hBN:C flakes, which exhibit homogeneous and reproducible optical emission. The thickness of the films, typically ranging between a few layers to a few tens of nanometers, was determined by optical force microscopy.\\

  For STM studies, hBN:C crystals were mechanically exfoliated onto 90 nm-thick Si/SiO$_2$ substrates. 1 up to 3 layer-thick films, identified via optical contrast and AFM techniques, were subsequently lifted with a PDMS/PC stamp and transferred onto a large graphite flake that was partially covered by a Cr/Au film, providing a conductive surface for the STM measurements. The samples were cleaned with DCM, ACE, and IPA, and finally annealed in an ultra-high vacuum in the STM chamber to remove any polymer residues from the transfer process. 

  \textbf{Optical characterisation} The PL spectra were measured in a back-scattering geometry under continuous-wave 514~nm excitation. The sample was cooled down via exchange gas in a closed-loop cryostat or via cold finger in a cryostat cooled with liquid helium. The laser was focused on the surface of the sample to a spot of about 1~$\upmu$m via an objective. The sample was positioned under the objective using \textit{x-y-z} piezo-scanner system. The PL signal was resolved by a spectrometer and detected by a charge-coupled device camera. The second-order photon correlations were measured in the Hanbury-Brown and Twiss configuration with avalanche photodiodes acting as photon detectors.

  \textbf{Scanning tunneling microscopy} The scanning tunneling microscopy was done in a low-temperature Createc system with base pressure below $10^{-10}$~mbar. In our samples, the tunnelling occurred between the tungsten tip and graphite substrate through a 3-layer-thick carbon-doped hBN barrier. The \textit{dI/dV} tunnelling spectra were measured at a modulated voltage between 3~meV and 10~meV at the frequency of 700-900~Hz. The tip was calibrated for spectroscopy against the surface state of gold in 111 orientation.

  \textbf{Theoretical methods} Density functional (DFT) electronic structure calculations are performed within the Vienna $ab$ $initio$ simulation package ({\sc vasp})~\cite{vasp1, vasp2} utilizing the projector-augmented wave (PAW)~\cite{Blochl2000} formalism with PBE  generalized-gradient approximation (GGA)~\cite{PBE} of exchange-correlation functional. For details on of the mapping to the minimal models and their solutions see SM.
    
\acknowledgements
This project was supported by the Ministry of Education (Singapore) through the Research Centre of Excellence program (grant EDUN C-33-18-279-V12, I-FIM), AcRF Tier 3 (MOE2018-T3-1-005). This material is based upon work supported by the Air Force Office of Scientific Research and the Office of Naval Research Global under award number FA8655-21-1-7026. J. Lu acknowledges the support from Agency for Science, Technology and Research (A*STAR) under its AME IRG Grant (Project No. M21K2c0113). K.W. and T.T. acknowledge support from JSPS KAKENHI (Grant Numbers 19H05790, 20H00354 and 21H05233). P.H. acknowledges the National Natural Science Foundation of China (51801041) and scholarship from the Guangxi Education Department (China). C.R.F acknowledges the European Union’s Horizon 2020 research and innovation programme under the Marie Skłodowska-Curie grant agreement N$^{\circ}$ 895369. M.P acknowledges the support from EU Graphene Flagship and FNP-Poland (IRA - MAB/2018/9 grant, SG 0P program of the EU). C.E.D. acknowledges support from the National Science Foundation under Grant No. DMR-2237674. D.I.B. was supported by the European Research Council (ERC) under the European Union’s Horizon 2020 research and innovation programme, grant agreement 854843-FASTCORR

\section*{Data availability}

The authors declare that the data supporting the findings of this study are available within the paper and in the supplementary material files.

\section{Supplemental Material}
\appendix
\section{Photoluminescence of \lowercase{h}BN in magnetic field}

\begin{figure}[b]
   \includegraphics[width=0.5\columnwidth]{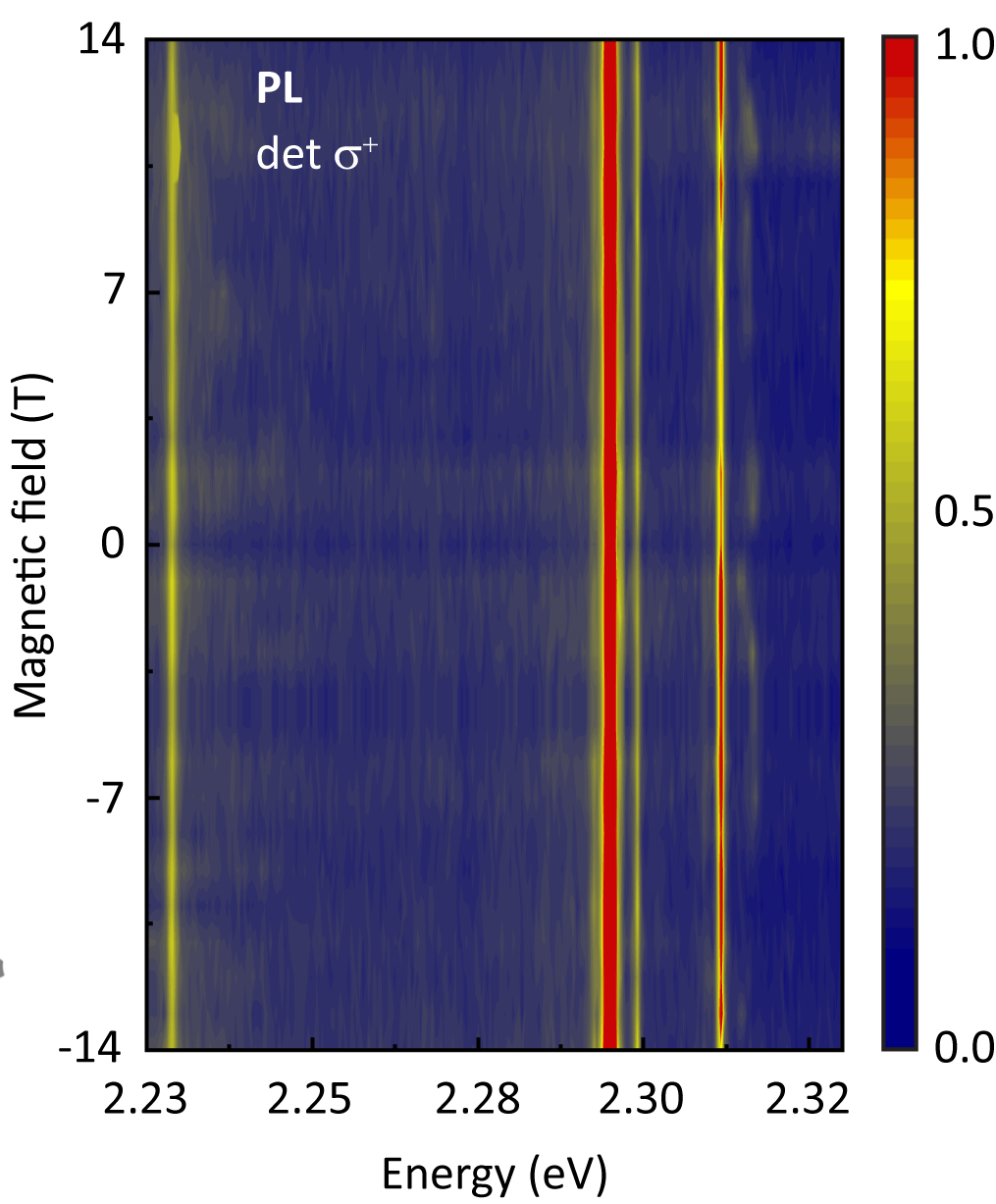}
    \caption{\label{fig:magnetoPL} Photoluminescence spectra of few layer hBN:C as a function of the magnetic field, showing the insensitivity of the transition energies. 
    }
\end{figure}

To gain a better sense of the nature of the observed spectral lines in photoluminescence (PL) experiments, we studied the optical response of the material in a magnetic field. Low-temperature micro-magneto-PL experiments were performed in the Faraday configuration, where the magnetic field is applied perpendicularly to the crystal plane. Measurements were carried out with the aid of a split-coil superconducting magnet up to 14~T. The sample was placed on top of an x-y-z piezo-stage kept at T = 4.2 ~K and was excited using a laser diode with energy below the hBN band gap. The combination of a quarter wave plate and a linear polarizer placed in the insert was used to analyse the circular polarization of the signal. The emitted light was dispersed with a 0.75 m long monochromator and detected with a charge-coupled device (CCD) camera. The result for $\sigma^{+}$ polarization from +14~T to -14~T (reversing the direction of the magnetic field yields the information corresponding to the other polarization component) is presented in Fig.\ref{fig:magnetoPL}. As can be appreciated from the false-colour map no signification changes can be observed while applying the external magnetic field, we do not detect any Zeeman splitting of the emission. The lack of magnetic field effects on the optical spectra of hBN:C is likely to indicate that the optically active transitions are dominated by singlet-singlet transitions, nevertheless, transitions between parallel branches of spin doublets and triplets cannot be ruled out.

\section{Theoretical methods  overview \label{sec:comp_details}}

    Density functional (DFT) electronic structure calculations are performed within the Vienna $ab$ $initio$ simulation package ({\sc vasp})~\cite{vasp1, vasp2} utilizing the projector-augmented wave (PAW)~\cite{Blochl2000} formalism with PBE  generalized-gradient approximation (GGA)~\cite{PBE} of exchange-correlation functional.  In these calculations, we set the energy cutoff 500~eV and the energy convergence criteria $10^{-8}$~eV and $3\times3$ $\Gamma$-centered k-mesh.  We start with the experimental bulk hBN lattice ($a$ = 2.51 \AA, $c$ = 6.66 \AA) using 5$\times$5$\times$1 unit cell size, where we remove or replace atomic positions with carbon atoms. All atomic positions are relaxed until all residual force components of each atom were less than $0.5 \times 10 ^{-3}$~eV/\AA.  To study the trends between bulk to thinner samples,  we construct defects embedded in an hBN monolayer with 15 \AA \, vacuum space along $z$ direction from the fully relaxed bulk geometries. This corresponds to the constrained monolayer case, while further relaxation of these supercells results in the free-standing monolayer setups. Excited states were studied by performing structural optimization with the constrained occupation of single-particle states (constrained  DFT).

\begin{figure}[h]
       \includegraphics[width=0.9\columnwidth]{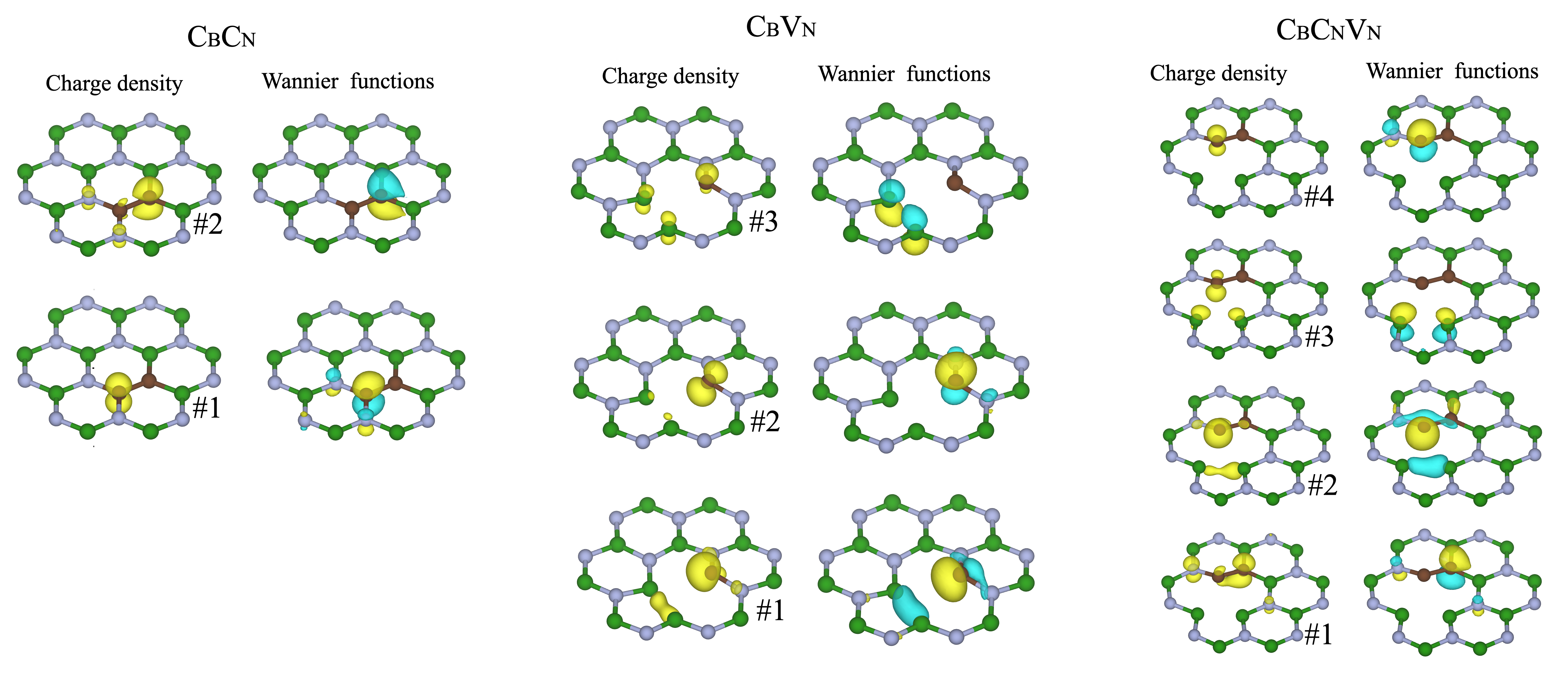}
        \caption{\label{fig:WF} Charge densities of defect-related states and constructed localized Wannier functions for defect complexes.  The leading atomic contributions of the partial charge densities were used as initial projections for the construction of Wannier functions (see Table~\ref{tab:WF}).
        }
    \end{figure}

    The defect related electronic states were constructed from Wannier functions~\cite{Marzari1997} (see Fig.~\ref{fig:WF}) $\phi_i$, which we calculate using the {\sc wannier90} package~\cite{Pizzi2020}. These Wannier functions form the basis to construct the extended Hubbard models. In order to preserve the symmetry of the system, we do not perform a maximal localization of the Wannier functions. This procedure allows to directly determine all hopping parameters $t_{ij} = \braket{\phi_i \vert H_{DFT} \vert \phi_j}$. Coulomb interaction matrix elements are calculated according to:
    \begin{align}
        U_{ijkl}&= \braket{\phi_i  \phi_j \vert \hat{U} \vert \phi_k  \phi_l } \\
                &= \int \int d^3r \, d^3r' \, \phi_i^*(r) \phi_k(r) \, U(r,r') \, \phi^*_j(r') \phi_l(r') \notag,
    \end{align}
    whereby we partially screen the interaction matrix elements within the constrained random phase approximation (cRPA) in the static limit ($\omega = 0$)~\cite{cRPA} using its recent implementation in {\sc VASP}~\cite{kaltak_merging_2015}. Additionally, to avoid double counting of Coulomb interaction terms acting on the single-particle (hopping) terms as a result of the DFT starting point, we use a Hartree double counting (DC) correction term~\cite{Muechler2021} of the form:
    \begin{align}
     H_{\text{DC}} = \sum_{i,j,\sigma} c^\dagger_{i\sigma}c_{j\sigma} \sum_{kl} P_{kl} U_{iljk},
    \end{align}
    where $P_{kl}$ is the DFT single particle density matrix in Wannier basis.  The chemical potential $\mu$ was chosen to constrain the nominal occupation of the defect states. All many-body states are derived from the exact diagonalization of the full Hamiltonian using the {\sc triqs} library~\cite{Parcollet2015}. 
    
\vspace{-0.1cm}
    
\newpage

    \begin{table}
    \caption{Atomic contributions to defect-related states in Fig.~\ref{fig:WF} \label{tab:WF}}
    \begin{ruledtabular}
    \begin {tabular}{c|c|c}
    \multirow{2}{*}{\cb{}\cn{}}  & \#1 & $0.07 \ket{C_1(p_z)} + 0.31 \ket{C_2(p_z)} $ \\
                                 & \#2 & $0.35 \ket{C_1(p_z)} + 0.04 \ket{C_2(p_z)} $ \\
    \hline  
    \multirow{3}{*}{\cb{}\vn{}}  & \#1 & $0.24\ket{C(p_x)}  + 0.06\ket{C(p_z)} $ \\
                                 & \#2 & $0.15 \ket{C(p_x)} + 0.33\ket{C(p_z)}$ \\ 
                                 & \#3 & $0.47\ket{B_c(p_z)}$  \\ 
    \hline  
    \multirow{4}{*}{\cb{}\cn{}\vn{}}  & \#1 & $0.08\ket{C_1(p_z)} + 0.26\ket{C_2(p_z)} $ \\
                                 & \#2 & $0.30\ket{C_1(p_x)}$ \\
                                 & \#3 & $0.12\ket{C_1{p_x}} + 0.10\ket{C_1{p_z}} + 0.20\ket{B_c(p_z)}$ \\
                                 & \#4 & $0.22\ket{C_1(p_z)} + 0.02 \ket{C_2(p_z)}$ \\
    \end {tabular}  
    \end{ruledtabular}
    \end {table}

\subsection{Details of Hamiltonian}

\begin{adjustbox}{varwidth=\textwidth,raise=9cm}
     In this section we provide single-particle hopping terms $t_{ij}$ (Table~\ref{tab:Hoppings}), density-density Coulomb and  exchange interactions (both bare in Table \ref{tab:Bare_Coulomb} and screened in Table~\ref{tab:Screened_Coulomb}) in the Wannier basis for defect complexes embedded in free-standing monolayer, constrained monolayer and bulk hBN lattices. A summary of orbitally averaged parameters is given in Table~\ref{tab:Averaged_parameters}.
\end{adjustbox}

\begin{table}[H]
\centering
\caption { Hopping matrices (in meV) }
\begin{ruledtabular}
\begin {tabular}{c|c|c|c}
  &  \cb{}\cn{}  & \cb{}\vn{}  & \cb{}\cn{}\vn{} \\
  \hline
 basis & $\ket{C_1(p_z) \, C_2(p_z)}$ &  $\ket{C(p_x) \, C(p_z) \, B_c(p_z)}$ & $\ket{C_1(p_z) \, C_2(p_z) \, C_1(p_x) \, B_c(p_z)}$ \\ 
  \hline
monolayer &
$\begin{pmatrix}
 2.158 & -1.673 \\
-1.673 &  0.930 \\
\end{pmatrix}$   
&
$\begin{pmatrix}
 0.516 & 0.756  &  0.606 \\
 0.756 & 1.221  & -0.469  \\
 0.606 &-0.469  &  1.753  \\
\end{pmatrix}$  
&
$\begin{pmatrix}
 0.783 & -1.504 &   0.388 &  -0.514  \\
-1.504 & -1.204 &  -0.023 &   0.181  \\
 0.388 & -0.023 &   0.387 &   0.494  \\
-0.514 &  0.181 &   0.494 &   1.328  \\
\end{pmatrix}$ \\
\hline 
constr. monolayer &
$\begin{pmatrix}
2.158  & -1.672 \\
-1.672 &  0.922 \\
\end{pmatrix}$   
&
$\begin{pmatrix}
0.404 & 0.708  &  0.473 \\
0.708 & 1.184  & -0.495  \\
0.473 & -0.495 & 1.618  \\
\end{pmatrix}$  
&
$\begin{pmatrix}
 0.688 & -1.443 &   0.341 &  -0.572  \\
-1.443 & -1.521 &  -0.087 &   0.098  \\
 0.341 & -0.087 &   0.213 &   0.212  \\
-0.572 &  0.098 &   0.212 &   1.126  \\
\end{pmatrix}$ \\
\hline
bulk &
$\begin{pmatrix}
2.517  & -1.315 \\
-1.315 &  0.277 \\
\end{pmatrix}$   
&
$\begin{pmatrix}
0.315  &   0.666 &  0.474  \\
0.666  &   1.145 & -0.446  \\
0.474  &  -0.446 &  1.521  \\
\end{pmatrix}$  
&
$\begin{pmatrix}
 0.658 & -1.382 &  0.323 &  -0.494  \\
-1.382 & -1.502 & -0.086 &   0.072  \\
 0.323 & -0.086 &  0.201 &   0.217  \\
-0.494 &  0.072 &  0.217 &   1.112  \\
\end{pmatrix}$
\end {tabular}
\end{ruledtabular}
\label{tab:Hoppings}
\end {table}

\begin{table}[H]
\centering
\caption { Bare Coulomb $v$ and  exchange  $j_H$ interaction matrices (in eV) }
\begin{ruledtabular}
\begin {tabular}{c|c|c|c}
  &  \cb{}\cn{}  & \cb{}\vn{}  & \cb{}\cn{}\vn{} \\
  \hline
 basis & $\ket{C_1(p_z) \, C_2(p_z)}$ &  $\ket{C(p_x) \, C(p_z) \, B_c(p_z)}$ & $\ket{C_1(p_z) \, C_2(p_z) \, C_1(p_x) \, B_c(p_z)}$ \\ 
   \hline
$v$ monolayer &
$\begin{pmatrix}
6.689 & 4.747 \\
4.747 & 6.597 \\
\end{pmatrix}$   
&
$\begin{pmatrix}
10.269 & 7.760 & 5.434 \\
 7.760 & 8.263 & 3.797  \\
 5.434 & 3.797 & 5.955  \\
\end{pmatrix}$  
&
$\begin{pmatrix}
6.879 &  4.935 &  7.115  &   3.622  \\
4.935 &  6.167 &  5.310  &   3.331  \\
7.115 &  5.310 &  9.822  &   5.186  \\
3.622 &  3.331 &  5.186  &   5.792  \\
\end{pmatrix}$ \\
\hline
$v$ constrained monolayer &
$\begin{pmatrix}
6.675 & 4.745 \\
4.745 & 6.603 \\
\end{pmatrix}$   
&
$\begin{pmatrix}
10.296 & 7.660 &  5.513 \\
7.660  & 8.052 &  3.806  \\
5.513  & 3.806 &  5.970  \\
\end{pmatrix}$  
&
$\begin{pmatrix}
6.585 &  4.885 &  6.966  &   3.576  \\
4.885 &  6.180 &  5.332  &   3.316  \\
6.966 &  5.332 &  9.877  &   5.200  \\
3.576 &  3.316 &  5.200  &   5.772  \\
\end{pmatrix}$ \\
\hline
$v$ bulk &
$\begin{pmatrix}
6.012 & 4.570 \\
4.570 & 6.270 \\
\end{pmatrix}$   
&
$\begin{pmatrix}
10.117 &  7.405 & 5.238  \\
 7.405 &  7.613 & 3.563  \\
 5.238 &  3.563 & 5.491  \\
\end{pmatrix}$  
&
$\begin{pmatrix}
5.877 &  4.528 &  6.569 &  3.302  \\
4.528 &  5.811 &  5.176 &  3.096  \\
6.569 &  5.176 &  9.830 &  4.938  \\
3.302 &  3.096 &  4.938 &  5.159  \\
\end{pmatrix}$  \\
\hline
$j_H$ monolayer &
$\begin{pmatrix}
6.689 &  0.155 \\
0.155 &  6.597 \\
\end{pmatrix}$   
&
$\begin{pmatrix}
10.269 & 0.461 & 0.174   \\
 0.461 & 8.263 & 0.039   \\
 0.174 & 0.039 & 5.955  \\
\end{pmatrix}$  
&
$\begin{pmatrix}
6.879 &  0.253 &  0.345  &  0.052  \\
0.253 &  6.167 &  0.071  &  0.055  \\
0.345 &  0.071 &  9.822  &  0.165   \\
0.052 &  0.055 &  0.165  &  5.792  \\
\end{pmatrix}$ \\
\hline
$j_H$ constrained monolayer &
$\begin{pmatrix}
6.675 & 0.155 \\
0.155 & 6.603 \\
\end{pmatrix}$   
&
$\begin{pmatrix}
10.296 & 0.456 &  0.184   \\
 0.456 & 8.053 &  0.040   \\
 0.184 & 0.040 &  5.970   \\
\end{pmatrix}$  
&
$\begin{pmatrix}
6.585 &  0.273 &  0.338  &  0.063  \\
0.273 &  6.181 &  0.063  &  0.057  \\
0.338 &  0.063 &  9.877  &  0.162  \\
0.063 &  0.057 &  0.162  &  5.773  \\
\end{pmatrix}$ \\
\hline
$j_H$ bulk &
$\begin{pmatrix}
6.012 & 0.202 \\
0.202 & 6.270 \\
\end{pmatrix}$   
&
$\begin{pmatrix}
10.117 &  0.446 & 0.173  \\
 0.446 &  7.613 & 0.040  \\
 0.173 &  0.040 & 5.491  \\
\end{pmatrix}$  
&
$\begin{pmatrix}
5.877 &  0.233 &  0.317 &  0.086  \\
0.233 &  5.812 &  0.060 &  0.053  \\
0.317 &  0.060 &  9.830 &  0.152  \\
0.086 &  0.053 &  0.152 &  5.160  \\
\end{pmatrix}$
\end {tabular}
\end{ruledtabular}
\label{tab:Bare_Coulomb}
\end {table}

\begin{table}[H]
\centering
\caption { Screened Coulomb $U$ and exchange $J_H$ interaction matrices (in eV) }
\begin{ruledtabular}
\begin {tabular}{c|c|c|c}
  &  \cb{}\cn{}  & \cb{}\vn{}  & \cb{}\cn{}\vn{} \\
  \hline
 basis & $\ket{C_1(p_z) \, C_2(p_z)}$ &  $\ket{C(p_x) \, C(p_z) \, B_c(p_z)}$ & $\ket{C_1(p_z) \, C_2(p_z) \, C_1(p_x) \, B_c(p_z)}$ \\ 
  \hline
$U$  monolayer &
$\begin{pmatrix}
3.238 & 2.504 \\
2.504 & 3.271 \\
\end{pmatrix}$   
&
$\begin{pmatrix}
4.777 & 3.428  &  2.741 \\
3.428 & 3.904  &  2.078  \\
2.741 & 2.078  &  3.045  \\
\end{pmatrix}$  
&
$\begin{pmatrix}
3.216 & 2.536  &  3.056  &  2.041  \\
2.536 & 2.982  &  2.622  &  1.967  \\
3.056 & 2.622  &  4.361  &  2.623  \\
2.041 & 1.967  &  2.623  &  2.874  \\
\end{pmatrix}$ \\
\hline

$U$  constrained monolayer &
$\begin{pmatrix}
3.233 & 2.503 \\
2.503 & 3.274 \\
\end{pmatrix}$   
&
$\begin{pmatrix}
4.765 & 3.378  &  2.757 \\
3.378 & 3.796  &  2.077  \\
2.757 & 2.077  &  3.044  \\
\end{pmatrix}$  
&
$\begin{pmatrix}
3.109 & 2.516  &  3.013  &  2.019  \\
2.516 & 2.995  &  2.629  &  1.959  \\
3.013 & 2.629  &  4.392  &  2.622  \\
2.019 & 1.959  &  2.622  &  2.872  \\
\end{pmatrix}$ \\
\hline

$U$ bulk &
$\begin{pmatrix}
1.903 &  1.424 \\
1.424 &  2.006 \\
\end{pmatrix}$   
&
$\begin{pmatrix}
3.423 & 2.144 & 1.587  \\
2.144 & 2.410 & 1.038  \\
1.587 & 1.038 & 1.741  \\
\end{pmatrix}$  
&
$\begin{pmatrix}
1.783 &  1.374 & 1.815 &  0.980  \\
1.374 &  1.788 & 1.539 &  0.950  \\
1.815 &  1.539 & 3.167 &  1.486  \\
0.980 &  0.950 & 1.486 &  1.576  \\
\end{pmatrix}$ \\
\hline

$J_H$  monolayer &
$\begin{pmatrix}
3.238 & 0.078 \\
0.078 & 3.271 \\
\end{pmatrix}$   
&
$\begin{pmatrix}
4.777 & 0.326  & 0.108  \\
0.326 & 3.904  & 0.020  \\
0.108 & 0.020  & 3.045  \\
\end{pmatrix}$  
&
$\begin{pmatrix}
3.216 & 0.115  &  0.241  &  0.025  \\
0.115 & 2.982  &  0.050  &  0.024  \\
0.241 & 0.050  &  4.361  &  0.098  \\
0.025 & 0.024  &  0.098  &  2.874  \\
\end{pmatrix}$ \\
\hline
$J_H$  constrained monolayer &
$\begin{pmatrix}
3.233 & 0.078 \\
0.078 & 3.274 \\
\end{pmatrix}$   
&
$\begin{pmatrix}
4.765 & 0.317  & 0.114  \\
0.317 & 3.797  & 0.021  \\
0.114 & 0.021  & 3.044  \\
\end{pmatrix}$  
&
$\begin{pmatrix}
3.109 & 0.121  &  0.234  &  0.029  \\
0.121 & 2.995  &  0.046  &  0.025  \\
0.234 & 0.046  &  4.393  &  0.097  \\
0.029 & 0.025  &  0.097  &  2.872  \\
\end{pmatrix}$ \\
\hline
$J_H$ bulk &
$\begin{pmatrix}
1.903 &  0.080 \\
0.080 &  2.006 \\
\end{pmatrix}$   
&
$\begin{pmatrix}
3.423 & 0.282 & 0.097  \\
0.282 & 2.411 & 0.017  \\
0.097 & 0.017 & 1.742  \\
\end{pmatrix}$  
&
$\begin{pmatrix}
1.783 &  0.089 & 0.202 &  0.031  \\
0.089 &  1.789 & 0.041 &  0.020  \\
0.202 &  0.041 & 3.168 &  0.084   \\
0.031 &  0.020 & 0.084 &   1.576  \\
\end{pmatrix}$ \\
\end {tabular} 
\end{ruledtabular}
\label{tab:Screened_Coulomb}
\end {table}

\begin{sidewaystable}
  \begin{table}[H]
        \caption{Comparison of  single-particle eigenvalues $\Delta E$, averaged values of bare intraorbital Coulomb $v_{0}$, screened intraorbital Coulomb $U_{0}$, screened interorbital Coulomb $U_{01}$, screened exchange  $J$ parameters (all in eV) and effective screening $\varepsilon$ for impurity complexes. \label{tab:Averaged_parameters}}
        \begin{ruledtabular}
            \begin{tabular}{c|ccc|ccc|ccc} 
                & \multicolumn{3}{c|}{\cb{}\cn{}} &    \multicolumn{3}{c|}{\cb{}\vn{}}  &  \multicolumn{3}{c}{\cb{}\cn{}\vn{}} \\
          \hline
                             & monolayer &  constr. monolayer  &  bulk  & monolayer  &   constr. monolayer   &  bulk  & monolayer  &   constr. monolayer   &  bulk \\  
          \hline
          $\Delta E$         & 3.56      & 3.56        &  3.45  & 1.95, 2.31  & 1.76, 2.19  &  1.72, 2.09  & 2.03, 3.34, 4.05         &    2.25, 3.14, 4.10 &  2.19, 3.11, 3.91  \\
          $v_{0}$            & 6.64      & 6.64        &  6.14  &  8.16       &    8.11     &     7.74     & 7.16     &           7.10      &       6.67         \\
          $v_{1}$            & 4.75      & 4.75        &  4.57  &  5.66       &    5.66     &     5.40     & 4.92     &           4.88      &       4.60         \\
          $j$                & 0.16      & 0.16        &  0.20  &  0.22       &    0.23     &     0.22     & 0.16     &           0.16      &       0.15         \\
          $U_{0}$            & 3.25      & 3.25        &  1.96  &  3.90       &    3.87     &     2.53     & 3.36     &           3.34      &       2.08         \\
          $U_{1}$            & 2.50      & 2.50        &  1.42  &  2.75       &    2.74     &     1.59     & 2.47     &           2.46      &       1.36         \\
          $J$                & 0.08      & 0.08        &  0.08  &  0.15       &    0.15     &     0.13     & 0.09     &           0.09      &       0.08         \\
          $\varepsilon$      & 2.0       & 2.0         &   3.2  &   2.1       &    2.1      &     3.2      & 2.1      &            2.1      &       3.3          \\
             \end{tabular}
        \end{ruledtabular}
    \end{table} 
\end{sidewaystable}

\subsection{\label{sec:DFT} Details of many-body spectra }

In this section we provide all many-body states for \cb{}\cn{} (Tables \ref{tab:MB_CBCN} and \ref{tab:MB_CBCN_orbital}), \cb{}\vn{} (Tables \ref{tab:MB_CBVN} and \ref{tab:MB_CBVN_orbital}) and \cb{}\cn{}\vn{} (Tables \ref{tab:MB_CBCNVN} and \ref{tab:MB_CBCNVN_orbital}) embedded into free-standing monolayer, constrained monolayer and bulk lattices of hBN.

\begin{table}[H]
\centering
\caption { Many-body spectra of \cb{}\cn{} (in band basis $\ket{\overline{b_1 b_2}; b_1 b_2}$, overline represents spin-down part). Only leading terms of eigenstates $\ket{\Psi_n}$ are shown.}
\begin{ruledtabular}
\begin {tabular}{c|c|c|c|c|c}
$E_n$ monolayer   &  $E_n$ constr. monolayer   & $E_n$ bulk     & $\ket{\Psi_n}$ & $S(S+1)$  & $m_s$ \\
\hline
0 &        0      &     0  &  $\ket{10;10}$    & 0  & 0     \\ 
\hline
\multirow{3}{*}{3.530}   &  \multirow{3}{*}{3.538} & \multirow{3}{*}{3.502} &$0.71\ket{01;10} - 0.71 \ket{10;01}$ & 2 & 0 \\
                         &     &                        & $\ket{00;11}$                       & 2 & 1 \\
                         &     &                        & $\ket{11;00}$                       & 2 & -1 \\   
\hline
4.163   &   4.171  & 3.976  &  $0.71\ket{01;10} - 0.71 \ket{10;01}$ &   0  & 0  \\
\hline
8.237   &   8.254  & 7.913  &  $\ket{01;01}$   & 0  & 0  \\
\end {tabular}
\end{ruledtabular}
\label{tab:MB_CBCN}
\end {table}

\begin{sidewaystable}
\begin{table}[H]
\centering
\caption { Many-body spectra of \cb{}\cn{} (in orbital basis $\ket{\overline{C_2(p_z) C_1(p_z)}; C_2(p_z) C_1(p_z) }$, overline represents spin-down part). Only leading terms of \textit{bulk} eigenstates $\ket{\Psi_n}$ are shown.}
\begin{ruledtabular}
\begin {tabular}{c|c|c|c|c|c}
$E_n$ monolayer   &  $E_n$ constr. monolayer   & $E_n$ bulk     & $\ket{\Psi_n}$ & $S(S+1)$  & $m_s$ \\
\hline
0 &        0      &     0  &  $0.14\ket{01;01} + 0.38\ket{01;10} + 0.38\ket{10;01} + 0.83\ket{10;10}$    & 0  & 0     \\ 
\hline
\multirow{3}{*}{3.530}   &  \multirow{3}{*}{3.538} & \multirow{3}{*}{3.502} & $-0.71\ket{01;10} + 0.71\ket{10;01}$ & 2 & 0 \\
                         &     &                        & $\ket{00;11}$                       & 2 & 1 \\
                         &     &                        & $\ket{11;00}$                       & 2 & -1 \\   
\hline
4.163   &   4.171  & 3.976  &  $0.47\ket{01;01} + 0.50\ket{01;10} + 0.50\ket{10;01} - 0.53\ket{10;10}$ &   0  & 0  \\
\hline
8.237   &   8.254  & 7.913  &  $0.87\ket{01;01} - 0.33\ket{01;10} - 0.33\ket{10;01} + 0.16\ket{10;10}$   & 0  & 0  \\
\end {tabular}
\end{ruledtabular}
\label{tab:MB_CBCN_orbital}
\end {table}

\begin{table}[H]
\centering
\caption { Many-body spectra of \cb{}\vn{} (in band basis $\ket{\overline{b_1 b_2 b_3}; b_1 b_2 b_3}$, overline represents spin-down part). Only leading terms of \textit{bulk} eigenstates $\ket{\Psi_n}$ are shown.}
\begin{ruledtabular}
\begin {tabular}{c|c|c|c|c|c}
$E_n$ monolayer  & $E_n$ constr. monolayer   & $E_n$ bulk  & $\ket{\Psi_n}$ &  $S(S+1)$  &    $m_s$ \\
\hline
0   &   0    &     0    &  $0.98\ket{100;100}$    &  0   & 0  \\
\hline
\multirow{3}{*}{2.639} &  \multirow{3}{*}{2.542} & \multirow{3}{*}{2.458} &  $0.68\ket{100;010} - 0.68\ket{010;100}$ & 2 & 0 \\
 &  &   &   $0.96\ket{000;110}$ & 2  &  1 \\
 &   &   &   $0.96\ket{110;000}$ & 2  & -1 \\
\hline
 3.410    &    3.243 &  3.041   &  $0.68\ket{100;010} + 0.68\ket{010;100}$  & 0  &  0 \\
\hline
\multirow{3}{*}{3.659} &  \multirow{3}{*}{3.502} & \multirow{3}{*}{3.342} & $0.67\ket{100;001} - 0.67\ket{001;100}$ & 2  & 0 \\
 &      &     &  $0.95\ket{000;101}$ & 2  &  1  \\
 &      &     &  $0.95\ket{101;000}$ & 2  & -1 \\ 
\hline
 4.022  &  3.900 &  3.662   &  $0.68\ket{100;001} + 0.68\ket{001;100}$   &  0  & 0 \\
\end {tabular}  
\end{ruledtabular}
\label{tab:MB_CBVN}
\end{table}
\end{sidewaystable}

\begin{sidewaystable}
\begin{table}[H]
\centering
\caption { Many-body spectra of \cb{}\vn{} (in orbital basis $\ket{\overline{B_c(p_z) C(p_z) C(p_x)}; B_c(p_z) C(p_z) C(p_x)}$, overline represents spin-down part). Only leading terms of \textit{bulk} eigenstates $\ket{\Psi_n}$ are shown.}
\begin{ruledtabular}
\begin {tabular}{c|c|c|c|c|c}
$E_n$ monolayer  & $E_n$ constr. monolayer   & $E_n$ bulk  & $\ket{\Psi_n}$ &  $S(S+1)$  &    $m_s$ \\
\hline
0   &   0    &     0    &  $0.74\ket{001;001} - 0.40\ket{001;010} - 0.40\ket{010;001}$    &  0   & 0  \\
\hline
\multirow{3}{*}{2.639} &  \multirow{3}{*}{2.542} & \multirow{3}{*}{2.458} &  $0.68\ket{001;010}  - 0.68\ket{010;001}$ & 2 & 0 \\
 &  &   &   $0.95\ket{000;011}$ & 2  &  1 \\
 &   &   &   $0.95\ket{011;000}$ & 2  & -1 \\
\hline
 3.410    &    3.243 &  3.041   &  $0.62\ket{001;001} - 0.34\ket{001;010}  - 0.34\ket{010;001} + 0.47\ket{010;010}$  & 0  &  0 \\
\hline
\multirow{3}{*}{3.659} &  \multirow{3}{*}{3.502} & \multirow{3}{*}{3.342} & $0.60\ket{001;100} -  0.60\ket{100;001}$ & 2  & 0 \\
 &      &     &  $-0.22\ket{000;011} + 0.85\ket{000;101} - 0.48\ket{000;110}$ & 2  &  1  \\
 &      &     &  $-0.22\ket{011;000} + 0.85\ket{101;000} - 0.48\ket{110;000}$ & 2  & -1 \\ 
\hline
 4.022  &  3.900 &  3.662   &  $0.57\ket{001;100} + 0.57\ket{100;001}$   &  0  & 0 \\
\end {tabular}  
\end{ruledtabular}
\label{tab:MB_CBVN_orbital}
\end {table}

\begin{table}[H]
\centering
\caption { Many-body spectra of \cb{}\cn{}\vn{} (in band basis $\ket{\overline{b_1 b_2 b_3 b_4}; b_1 b_2 b_3 b_4}$, overline represents spin-down part). Only leading terms of  \textit{bulk} eigenstates $\ket{\Psi_n}$ are shown.}
\begin{ruledtabular}
\begin {tabular}{c|c|c|c|c|c}
$E_n$ monolayer  &  $E_n$ constr. monolayer  & $E_n$ bulk  & $\ket{\Psi_n}$ &   $S(S+1)$   & $m_s$ \\
\hline
0     &   0     &     0    &  $0.98\ket{1000;1100}$ &  0.75  &  0.5   \\
0     &   0     &     0    &  $0.98 \ket{1100;100}$ &  0.75  & -0.5   \\
\hline
2.315 &    2.076 &  2.049   &  $0.97\ket{1000;1010}$ &  0.75  &  0.5   \\
2.315 &    2.076 &  2.049   &  $0.97\ket{1010;1000}$ &  0.75  & -0.5   \\
\hline
2.300 &    2.525 &  2.400   &  $0.99\ket{0100;1100}$ &  0.75 &  0.5   \\
2.300 &    2.525 &  2.400   &  $0.99\ket{1100;0100}$ &  0.75 & -0.5   \\
\hline
\multirow{4}{*}{3.340} &  \multirow{4}{*}{3.279}  &  \multirow{4}{*}{3.220}    &  $0.54\ket{1000;0110} - 0.54\ket{0100;1010} + 0.54\ket{0010;1100}$ &   3.75  &  0.5  \\ 
  &          &            &  $0.54\ket{1100;0010} - 0.54\ket{1010;0100} + 0.54\ket{0110;1000}$ &   3.75  & -0.5  \\
  &          &            &  $0.94 \ket{0000;1110}$ &  3.75  &   1.5 \\
  &          &            &  $0.94\ket{1110;0000}$ &  3.75  &  -1.5 \\
\hline      
3.308  &    3.335 & 3.252    &   $0.98\ket{1000;1001}$  & 0.75   &  0.5  \\    
3.308  &    3.335 & 3.252    &   $0.98\ket{1001;1000}$  & 0.75   & -0.5  \\    
\end {tabular}  
\end{ruledtabular}
\label{tab:MB_CBCNVN}
\end{table}
\end{sidewaystable}

\begin{sidewaystable}

\begin{table}[H]
\centering
\caption { Many-body spectra of \cb{}\cn{}\vn{} (in orbital basis $\ket{\overline{B_c(p_z) C_1(p_x) C_2(p_z) C_1(p_z)}; B_c(p_z) C_1(p_x) C_2(p_z) C_1(p_z)}$, overline represents spin-down part). Only leading terms of \textit{bulk} eigenstates $\ket{\Psi_n}$ are shown.}
\begin{ruledtabular}
\begin {tabular}{c|c|c|c|c|c}
$E_n$ monolayer  &  $E_n$ constr. monolayer  & $E_n$ bulk  & $\ket{\Psi_n}$ &   $S(S+1)$   & $m_s$ \\
\hline
0     &   0     &     0    &  $0.39\ket{0001;0110}  + 0.39\ket{0010;0101} + 0.75\ket{0010;0110}$ &  0.75  &  0.5   \\
0     &   0     &     0    &  $0.39\ket{0101;0010}  + 0.39\ket{0110;0001} + 0.75\ket{0110;0010}$ &  0.75  & -0.5   \\
\hline
2.315 &    2.076 &  2.049   &  $0.32\ket{0001;0011} + 0.59\ket{0010;0011} - 0.52\ket{0010;1010}$ &  0.75  &  0.5   \\
2.315 &    2.076 &  2.049   &  $0.32\ket{0011;0001} + 0.59\ket{0011;0010} - 0.52\ket{1010;0010}$ &  0.75  & -0.5   \\
\hline
2.300 &    2.525 &  2.400   &  $  0.35\ket{0100;0011} + 0.33\ket{0100;0101} + 0.74\ket{0100;0110}$ &  0.75 &  0.5   \\
2.300 &    2.525 &  2.400   &  $  0.34\ket{0011;0100} + 0.33\ket{0101;0100} + 0.74\ket{0110;0100}$ &  0.75 & -0.5   \\
\hline
\multirow{4}{*}{3.340} &  \multirow{4}{*}{3.279}  &  \multirow{4}{*}{3.220}    &  $-0.48\ket{0011;0100} + 0.48\ket{0101;0010} - 0.48\ket{0110;0001}$ &   3.75  &  -0.5  \\ 
  &          &            &  $-0.48\ket{0001;0110}  + 0.48\ket{0010;0101} - 0.48\ket{0100;0011}$ &   3.75  & 0.5  \\
  &          &            &  $0.83\ket{0000;0111} + 0.23\ket{0000;1101} + 0.52\ket{0000;1110}$ &  3.75  &   1.5 \\
  &          &            &  $0.83\ket{0111;0000} + 0.23\ket{1101;0000} + 0.52\ket{1110;0000}$ &  3.75  &  -1.5 \\
\hline      
3.308  &    3.335 & 3.252    &   $ 0.59\ket{0010;0011} + 0.53\ket{0010;1010} $  & 0.75   &  0.5  \\    
3.308  &    3.335 & 3.252    &   $ 0.59\ket{0011;0010} + 0.53\ket{1010;0010}$  & 0.75   & -0.5  \\    
\end {tabular}  
\end{ruledtabular}
\label{tab:MB_CBCNVN_orbital}
\end{table}

\end{sidewaystable}

\begin{table}[H]
    \caption{Evolution of  Huang-Rhys factor \shr{}  for impurity complexes embedded into free-standing monolayer, constrained monolayer, and bulk hBN. \label{tab:Huang-Rhys}}
    \begin{ruledtabular}
    \begin {tabular}{c|ccc|ccc|ccc}
  & \multicolumn{3}{c|}{\cb{}\cn{}} &   \multicolumn{3}{c|}{\cb{}\vn{}} &  \multicolumn{3}{c}{\cb{}\cn{}\vn{}}  \\
    \hline
    & monolayer & constr. monolayer  & bulk  & monolayer & constr. monolayer  & bulk  & monolayer & constr. monolayer  & bulk  \\
    \hline
1  &  0.65    &  0.63 &  0.75  & 33.0   &  18.2   &  17.2 &  24.8   &   12.6   &  12.2  \\  
2  &          &       &        &  8.0   &   3.8   &   3.7 &  15.8   &   12.9   &  12.8  \\ 
3  &          &       &        &        &         &       &  11.0   &    2.9   &   3.4  \\ 
    \end{tabular}  
    \end{ruledtabular}
    \end{table}


\bibliography{CHBn}


\end{document}